\documentclass[runningheads]{llncs}

\usepackage{graphicx}
\usepackage{etoolbox}
\makeatletter
\let\llncs@addcontentsline\addcontentsline
\patchcmd{\maketitle}{\addcontentsline}{\llncs@addcontentsline}{}{}
\patchcmd{\maketitle}{\addcontentsline}{\llncs@addcontentsline}{}{}
\patchcmd{\maketitle}{\addcontentsline}{\llncs@addcontentsline}{}{}
\makeatother
\usepackage{hyperref}
\usepackage{bookmark}
\hypersetup{pdfborder = {0 0 0}}

\usepackage{mathtools, amsmath, amssymb, colonequals}
\usepackage[algoruled, linesnumbered, noline]{algorithm2e}

\usepackage{ifsym}
\usepackage{wrapfig}
\usepackage{microtype}
\usepackage{mdframed}
\usepackage{color,soul}
\usepackage{multirow}
\usepackage{paralist}
\usepackage{float}
\usepackage{marvosym}
\usepackage{orcidlink}
\usepackage{tabularx}

\graphicspath{{figures/}{./figures/}}
\usepackage{tikz}
\usetikzlibrary{calc}

\newcommand{\goes}[1]{\xrightarrow{#1}}
\newcommand{\crn}{\mathcal{N}}

\newcommand{\N}{\mathbb{N}}

\newcommand{\pc}{c}
\newcommand{\pk}{k}

\usetikzlibrary{calc}

\definecolor{nicebg}{HTML}{f6f0e4}
\definecolor{nicered}{HTML}{7f0a13}
\definecolor{nicebgred}{HTML}{f2e7e8}
\definecolor{niceblue}{HTML}{104354}
\definecolor{nicebgblue}{HTML}{e8edee}
\definecolor{nicegreen}{HTML}{217516}
\definecolor{nicebggreen}{HTML}{e9f1e8}
\definecolor{nicepurple}{HTML}{884bab}
\definecolor{nicebgpurple}{HTML}{f3edf7}
\definecolor{niceorange}{HTML}{d27c11}
\definecolor{nicebgorange}{HTML}{fbf2e8}
\definecolor{nicepink}{HTML}{e95f9f}
\definecolor{nicebgpink}{HTML}{fdeff6}
\definecolor{niceredlight}{HTML}{c9888d}
\definecolor{nicebluelight}{HTML}{78a4b8}
\definecolor{nicegreenlight}{HTML}{76de68}
\definecolor{nicepurplelight}{HTML}{bc87db}
\definecolor{niceredbright}{HTML}{bd0310}
\definecolor{nicebgredbright}{HTML}{f9e6e8}
\definecolor{nicebluebright}{HTML}{197b9b}
\definecolor{nicebgbluebright}{HTML}{e8f2f5}

\newcommand{\applySegment}[5]{
	\foreach[count=\i] \o in #5
	{
		\pgfmathsetmacro{\iminuseins}{{int(\i-1)}}%
		\pgfmathsetmacro{\prev}{{\i==1 ? "#3" : "#3\iminuseins"}}%
		\node[#1] (#3\i) at ($ (\prev) + (\o)$) {};
		\draw[->, #2] (\prev) -- (#3\i);
		\node[#1] (#4) at ($ (\prev) + (\o)$) {};
	}
}

\newcommand{\drawAbstractState}[9]{
	\coordinate (#8_BL) at ($(#2) + (-0.5,-0.5)$);
	\coordinate (#8_BR) at ($ (#8_BL) + (#3,0)$);
	\coordinate (#8_TL) at ($ (#8_BL) + (0,#4)$);
	\coordinate (#8_TR) at ($ (#8_BL) + (#3,#4)$);
	\fill[#9] (#8_BL) rectangle (#8_TR);
	\draw[#1] (#8_BL) -- (#8_BR);
	\draw[#1] (#8_BL) -- (#8_TL);
	\draw[#1] (#8_BR) -- (#8_TR);
	\draw[#1] (#8_TL) -- (#8_TR);
	\draw[#1] (#8_BL) -- ($ (#8_BL) - (#5,0) $);
	\draw[#1] (#8_BL) -- ($ (#8_BL) - (0,#5) $);
	\draw[#1] (#8_BR) -- ($ (#8_BR) + (#5,0) $);
	\draw[#1] (#8_BR) -- ($ (#8_BR) - (0,#5) $);
	\draw[#1] (#8_TL) -- ($ (#8_TL) - (#5,0) $);
	\draw[#1] (#8_TL) -- ($ (#8_TL) + (0,#5) $);
	\draw[#1] (#8_TR) -- ($ (#8_TR) + (#5,0) $);
	\draw[#1] (#8_TR) -- ($ (#8_TR) + (0,#5) $);
	\node[#7] (#8) at  ($ (#8_BL) + (#3/2,#4/2) $) {#6};
}

\begin{document}
\title{
Abstraction-Based Segmental Simulation of Chemical Reaction Networks\thanks{This work has been supported by the Czech Science Foundation grant \mbox{GJ20-02328Y}, the German Research Foundation (DFG) projects 378803395 (ConVeY) and 427755713 (GOPro) as well as the ERC Advanced Grant 787367 (PaVeS).
}
}
\author{Martin Helfrich\inst{1}(\Letter) \orcidlink{0000-0002-3191-8098} \and Milan \v{C}e\v{s}ka\inst{2}(\Letter)\orcidlink{0000-0002-0300-9727} \and Jan K\v{r}et\'{i}nsk\'{y}\inst{1}\orcidlink{0000-0002-8122-2881}
 \and \v{S}tefan~Marti\v{c}ek\inst{2}\orcidlink{0000-0003-4498-7436} 
 \vspace{-1em}
%\thanks{}
}

\authorrunning{M. Helfrich et al.}
\institute{Technical University of Munich, Munich, Germany \\ \email{helfrich@in.tum.de}
\and Brno University of Technology, Brno, Czech Republic\\ \email{ceaskam@fit.vutbr.cz}}

\maketitle              % typeset the header of the contribution

\vspace{-1em}
\begin{abstract}
Simulating chemical reaction networks is often computationally demanding, in particular due to stiffness.
We propose a novel simulation scheme 
where long runs are not simulated as a whole but assembled from shorter precomputed segments of simulation runs.
On the one hand, this speeds up the simulation process to obtain multiple runs since we can reuse the segments.
On the other hand, questions on diversity and genuineness of our runs arise.
However, we ensure that we generate runs close to their true distribution by generating an appropriate abstraction of the original system and utilizing it in the simulation process. 
Interestingly, as a by-product, we also obtain a yet more efficient simulation scheme, yielding runs over the system's abstraction.
These provide a very faithful approximation of concrete runs on the desired level of granularity, at a low cost.
Our experiments demonstrate the speedups in the simulations while preserving key dynamical as well as quantitative properties. 

\end{abstract}

\newcommand{\para}[1]{\noindent\textbf{#1} }

\vspace{-1.5em}
\section{Introduction}

\para{Chemical Reaction Networks (CRNs)} are a versatile language widely used for {modeling and analysis} of biochemical systems~\cite{Chellaboina} as well as for high-level {programing} of molecular devices~\cite{soloveichik2010dna,cardelli2013two}. The time-evolution of CRNs is governed by the Chemical Master Equation that leads to a (potentially infinite) discrete-space, continuous-time Markov chain (CTMC) with ``population'' structure, describing how the probability of the molecular counts of each chemical species evolve in time.   
Many important biochemical systems feature complex dynamics, that are hard to analyze due to \emph{state-space explosion, stochasticity, stiffness,  and multimodality} of the population distributions~\cite{Kampen1992b,goutsias2005quasiequilibrium}.
This fundamentally limits the class of  systems the existing techniques can handle effectively. 
There are several classes of approaches that try to circumvent these issues, in particular, (i) \emph{stochastic simulation} avoids the explicit construction of the state space by sampling trajectories in the CRN, and (ii) \emph{abstraction} builds a smaller/simpler model preserving the key dynamical properties and allowing for an efficient numerical analysis of the original CRN. Over the last two decades, there has been very active research on improving the performance and precision of these approaches, see the related work below. 
Yet, running thousands of simulations to approximate the stochastic behavior often takes many hours; and abstractions course enough to be analyzed easily often fail to capture the complex dynamics, e.g. oscillations in the notorious tiny (two-species) predator-prey system.

\vspace{0.5em}
\para{Our contribution.} In this paper, in several simple steps, we uniquely \emph{combine} the simulation methods with the abstraction methods for CTMC states, further narrowing the performance gap. 
As the first step, we suggest leveraging memoization, a general optimization technique that pre-computes and stores partial results to speedup the consequent computation. In particular, when simulating from a current state we reuse previously generated pieces of runs, called \emph{segments}, that start in ``similar enough'' states. Thus, rather than spending time on simulating a whole new run, we quickly stitch together the segments. 
To ensure a high variety of runs and generally a good correspondence to
the original
probability space of runs, we not only have to generate a sufficiently
large number of segments;
but it is crucial to also consider their length and the similarity of
their starting states.
We show how the latter two questions can be easily answered using the standard \emph{interval abstraction} on the populations, yielding faithful yet fast simulations of the CTMC for the CRN.

In a second step, we also produce simulation runs over the (e.g. interval) abstraction of the CTMC, not only fast but also with low memory requirements, allowing for efficient analysis on the desired level of detail.
To this end, we drop all the concrete information of each segment, keeping only its abstraction plus its \emph{concrete end state}. 
This surprising choice of information allows us to \emph{define transitions on the abstraction using the simulation} in a rather non-standard way.
The resulting semantics and dynamics of the abstraction are non-Markovian, but capture the dynamics of the analyzed system very precisely.
From the methodological perspective, the most interesting point is that simulation and abstraction can \emph{help each other} although typically seen as disparate tools.
  
\vspace{-1em}
\subsubsection{Related work.}

To speed up the standard Stochastic Simulation Algorithm (SSA)~\cite{gillespie1977exact}, several 
\emph{approximate multi-scale} simulation techniques have been proposed. They include advanced $\tau$-leaping methods~\cite{cao2006efficient,lester2015adaptive}, that use a Poisson approximation to adaptively take time steps leaping over many reactions assuming the system does not change significantly within these steps. 
Alternatively, various partitioning schemes for fast and slow reactions have been considered~\cite{cao2005slow} allowing one to approximate the fast reactions by a quasi-steady-state assumption~\cite{rao2003stochastic,goutsias2005quasiequilibrium}.  The idea of separating the slow and fast sub-networks has been further elaborated in hybrid simulations treating some appropriate species as continuous variables and the others as discrete ones~\cite{salis2005accurate}. As before, appropriate partitioning of the species is essential for the performance and accuracy, and thus several (adaptive) strategies have been proposed~\cite{ganguly2015jump,hepp2015adaptive}. Recently a \emph{deep learning} paradigm has been introduced to further shift the scalability of the CRN analysis. In~\cite{cairoli2021abstraction}, the authors learn from a set of stochastic simulations of the CRN a generator, in the form of a Generative Adversarial Network, that efficiently produces trajectories having a similar distribution as the trajectories in the original CRN. In~\cite{gupta2021deepcme} the authors go even further and learn from the simulations an estimator of the given statistic over the original CRN. The principal limitation of these approaches is the overhead related to the learning phase that typically requires a nontrivial number of the simulations of the original CRN.

To build a plausible and computationally tractable abstraction of CRNs, various \emph{state-space reduction techniques} have been proposed that either truncate states of the underlying CTMC with insignificant probability~\cite{munsky2006finite,Henzinger2009,Mateescu10} or leverage structural properties of the CTMC to aggregate/lump selected sets of states~\cite{abate2021adaptive,backenkohler2021abstraction}.
The \emph{interval abstraction} of the species population is a widely used approach to mitigate the state-space explosion problem~\cite{Zhang09,ferm2009adaptive,Madsen2012}.
We define a segment of simulation runs as a sequence of transitions that can be seen as a single transition of the abstracted system.
These ``abstract'' transitions in the interval abstractions are studied in \cite{CAV19} as ``accelerated'' transitions.
Alternatively, several hybrid models have been considered levering a similar idea as the hybrid simulations. In~\cite{henzinger2010hybrid}, a pure deterministic semantic for large population species is used. The moment-based description for medium/high-copy number species was used in~\cite{Verena2013}. The LNA approximation and an adaptive partitioning of the species according to leap conditions (that is more general than partitioning based on population thresholds) were proposed in~\cite{HybridLNA2016}.  

\vspace{-1em}
\subsubsection{Advantages of our approach.}
We show that the proposed \emph{segmental simulation scheme} preserves the key dynamical properties and its qualitative accuracy (with respect to the SSA baseline) is comparable with advanced simulation as well as deep-learning approaches. The scheme, however, provides a significant computational gain over these approaches. Consider a detailed analysis (including 100000 simulation runs) of the famous Toggle switch model reported in~\cite{hepp2015adaptive}. Using the $\tau$-leaping implementation in StochPY~\cite{maarleveld2013stochpy} is not feasible and the state-of-the-art adaptive hybrid simulation method requires a day to perform such an analysis. However, our approach needs less than two hours. Moreover, our lazy strategy does not require a computationally demanding pre-computation and typically significant benefits from reusing segments after a small number of simulations. This is the key advantage compared to the learning approaches~\cite{cairoli2021abstraction,gupta2021deepcme}, where a large number of simulations of the original CRN are required, as well as to approaches based on (approximate) bisimulation/lumping~\cite{larsen1991bisimulation,desharnais2008approximate}, requiring a complex analysis of the original model. 
The approaches based on hybrid formal analysis of the underlying CTMC~\cite{henzinger2010hybrid,Verena2013,HybridLNA2016} have to perform 
a computationally  demanding analysis of conditioned stochastic processes. For example, in~\cite{HybridLNA2016} the authors report that an analysis of the Viral infection model took more than 1 hour. The segmental simulation using 10000 runs provides the same quantitative information in 3 minutes. 

In our previous work~\cite{CAV19}, we proposed a semi-quantitative abstraction and analysis of CRNs focusing on explainability of the results and low computational complexity, however, providing only limited quantitative accuracy. The proposed simulation scheme provides significantly better accuracy as it keeps track of the current concrete state and thus avoids ``jumps'' to the abstract state's representative. This kind of rounding is a major source of error as exemplified in the predator-prey model, where the semi-quantitative
abstraction of~\cite{CAV19} failed to accurately preserve the oscillation and our new approach captures it faithfully.

\vspace{-0.5em}
\section{Preliminaries} 

\subsection*{Chemical Reaction Networks}

\paragraph{CRN Syntax.}
A \emph{chemical reaction network (CRN)} $\crn=(\Lambda,\mathcal{R})$ is a pair of finite sets, where $\Lambda$ is a set of \emph{species}, $|\Lambda|$ denotes its size, and $\mathcal{R}$ is a set of reactions. Species in $\Lambda$ interact according to the reactions in $\mathcal{R}$. A \emph{reaction} $\tau \in \mathcal{R}$ is a triple $\tau=(r_{\tau},p_{\tau},k_{\tau})$, where $r_{\tau} \in  \mathbb{N}^{|\Lambda|}$ is the \emph{reactant complex}, 
$p_{\tau} \in  \mathbb{N}^{|\Lambda|}$ is the \emph{product complex} and $k_{\tau} \in \mathbb{R}_{>0} $ is the coefficient associated with the rate of the reaction. $r_{\tau}$ and $p_{\tau}$ represent the stoichiometry of reactants and products.
Given a reaction $\tau_1=(  [1,1,0],[0,0,2],k_1 )$, we often refer to it as $\tau_1 : \lambda_1 + \lambda_2 \, \goes{k_1}  \,    2\lambda_3 $. 
\vspace{-0.3em}
\paragraph{CRN semantics.}\label{Concrete Semantics}
Under the usual assumption of mass action kinetics\footnote{We can  handle alternative kinetics including Michaelis–Menten and Hill kinetics.}, the \emph{stochastic} semantics of a CRN $\crn$ is generally given in terms of a discrete-state, continu\-ous-time stochastic process $\mathbf{X(t)}=(X_1(t),X_2(t), \ldots, X_{|\Lambda|}(t) ,t\geq 0)$ \cite{ethier2009markov}.
The \emph{state change} associated with the reaction $\tau$ is defined by $\upsilon_{\tau}=p_{\tau} - r_{\tau}$, i.e. the state $\mathbf{X}$ is changed to $\mathbf{X}' = \mathbf{X} + \upsilon_{\tau}$. 
For example, for $\tau_1$ as above, we have $\upsilon_{\tau_1}=[-1,-1,2]$. 
A reaction can only happen in a state $\mathbf{X}$ if all reactants are present in sufficient numbers. Then we say that the reaction is enabled in $\mathbf{X}$.
The behavior of the
stochastic system $\mathbf{X(t)}$ can be described by the (possibly infinite) continuous-time Markov
chain (CTMC). The transition rate corresponding to a reaction $\tau$ is given by a \emph{propensity function} that in general depends on the stoichiometry of reactants, their populations and the coefficient~$k_{\tau}$.   

\vspace{-0.5em}
\subsection*{Related Concepts}

\paragraph{Population level abstraction.}
The CTMC is the accurate representation of CRN~$\crn$, but---even when finite--- it is not scalable in practice because of the state space explosion problem \cite{kwiatkowska2014probabilistic,heath2008probabilistic}. Various (adaptive) population abstractions~\cite{Zhang09,ferm2009adaptive,Madsen2012,abate2021adaptive} have been proposed to reduce the state-space and preserve the dynamics of the original CRN. Intuitively, \textit{abstract states} are given by intervals on sizes of populations (with an additional specific that the abstraction captures enabledness of reactions). In other words, the population abstraction divides the \textit{concrete states} of the CTMC into hyperrectangles called abstract states. We chose one concrete state within each abstract state as its \textit{representative}.
Although our approach is applicable to very general types of abstractions, for simplicity and specificity we consider in this paper only the \emph{exponential partitioning}
for some parameter $1 < \pc \leq 2$ given as $\{[0{,}0]\} \cup \{[\lfloor \pc^{n-1}\rceil, \lfloor \pc^{n} \rceil-1] \ :\ n \in \N\}$ for all dimensions. For example with $c{=}2$ the intervals are $[0{,}0], [1{,}1], [2{,}3], [4{,}7], [8{,}15], \dots$ i.e. they grow exponentially in~$\pc$.
While the structure of the abstract states is rather standard, the transitions between the abstract states are defined in different ways.

\paragraph{Stochastic simulation.} An alternative computational approach to the analysis of CRNs is to generate trajectories using stochastic algorithms for simulation. Gillespie's stochastic simulation algorithm (known as SSA)~\cite{gillespie1977exact} is a widely used exact version of such algorithms, which produces statistically correct trajectories, i.e., sampled according to the stochastic process described by the Chemical Master Equation. To produce such a trajectory, SSA repeatedly applies one reaction at a time while keeping track of the elapsed time. This can take a long time if the number of reactions per trajectory is large. 
This is typically the case if (1) there are large numbers of molecules, (2) the system is stiff (with high differences in rates) or (3) we want to simulate the system for a time that is long compared to the rates of a single reaction. One of the approaches that mitigate the efficiency problem is $\tau$-leaping~\cite{gillespie2001approximate}. The main idea is that for a given time interval (of length~$\tau$), where the reaction propensities do not change significantly, it is sufficient to sample (using Poisson distributions) only the number of occurrences for each reaction and not their concrete sequence. Having the numbers allows one to compute and apply the joint effect of the reactions at once. As detailed in the next section, instead of this time locality, we leverage a space locality.

\section{The plan: A technical overview}
Since we shall work with different types of simulation runs, gradually building on top of each other, and both with the concrete system and its abstraction, we take the time here to overview the train of thoughts, the involved objects, and the four main conceptual steps:
\begin{itemize}
    \item Section \ref{ssec:segments} introduces \textbf{segmental simulation} as a means to obtain simulation runs of the concrete system faster at the cost of (i) a significant memory overhead and (ii) skewing the probability space of the concrete runs, but only negligibly w.r.t. a user-given abstraction of the state space.
    \item Section \ref{ssec:densely} introduces \textbf{densely concrete simulation}, which eliminates the memory overhead but produces concrete simulation runs where only some of the concrete states on the run are known, however, frequently enough to get the full picture (again w.r.t. the abstraction), see Fig. \ref{fig:pred_prey_examples} (bottom).
    \item Section \ref{sec:4.1} shows how to utilize Section \ref{sec:segmental} to equip the state-space abstraction (quotient) with a powerful transition function and semantics in terms of a probability space over abstract runs, i.e. runs over the abstract state space.
    The \textbf{abstraction is executable} and generates \textbf{abstract simulation runs} with extremely low memory requirements, yet allowing for transient analyses that are very precise (again w.r.t. the abstraction), see Fig. \ref{fig:abstract} (left).
    \item Section \ref{sec:4.2} considers the \textbf{concretization of abstract simulation runs} 
    back to the concrete space, see Fig. \ref{fig:abstract} (right). 
\end{itemize}

\section{Segmental Simulation via Abstract States} \label{sec:segmental}

\subsection{Computing and Assembling Segments via Abstract States} \label{ssec:segments}

\subsubsection{Precomputing segments.}  
Assume we precomputed for each concrete state $s$ a list of $\pk$ randomly chosen short trajectories, called \textit{segments}, starting in $s$.
Note that we may precompute multiple segments with the same endpoint, reflecting that this evolution of the state is more probable.
We can now obtain a trajectory of the system by repeatedly sampling and applying a precomputed segment for the current state instead of a single reaction.

\vspace{-0.5em}
\subsubsection{Using abstraction.}
While simulating with already precomputed segments would be faster, it is obviously inefficient to precompute and store the segments for each state separately.
However, note that the rates of reactions in CRNs are similar for states with similar amounts of each species, in particular for states within the same abstract state of the population-level abstraction.
Consequently, we only precompute $\pk$ segments for one concrete state per abstract state: the abstract state's representative (typically its center).
For other states, the distribution of the segments would be similar and our approximation assumes them to be the very same. 
While the exponential population-level abstraction is a good starting point for many contexts, the user is free to provide any partitioning (quotient) of the state space that fits the situation and the desired granularity of the properties in question. E.g.~one could increase the number of abstract states in regions of the state space we want to study.

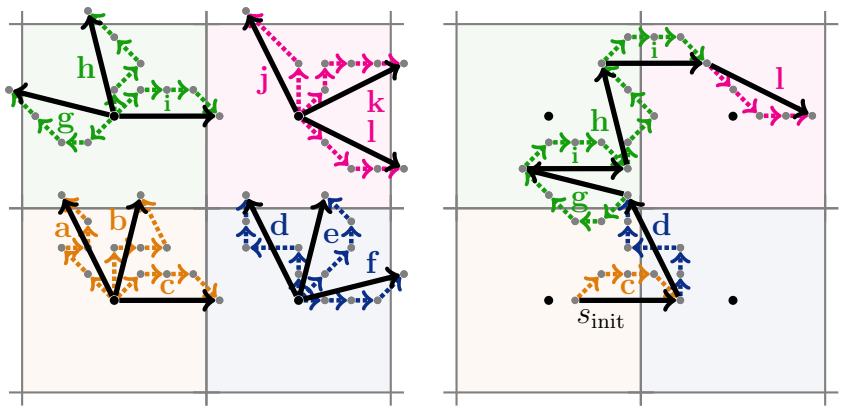
\begin{figure}[!htb]
    \centering
    \begin{tikzpicture}[scale=0.35, auto]
	\newcommand{\segA}{{-1,1},{-1,1},{1,0},{0,1},{-1,1}}
	\newcommand{\segB}{{0,2},{1,0},{1,0},{-1,2}}
	\newcommand{\segC}{{1,1},{1,0},{1,0},{1,-1}}
	
	\newcommand{\segD}{{0,1},{0,1},{-2,0},{0,1},{0,1}}
	\newcommand{\segE}{{1,1},{1,1},{0,1},{-1,1}}
	\newcommand{\segF}{{1,0},{1,0},{1,0},{1,1}}
	
	\newcommand{\segG}{{-1,-1},{-1,0},{-1,1},{-1,1}}
	\newcommand{\segH}{{0,1},{1,1},{-1,1},{-1,1}}
	\newcommand{\segI}{{1,1},{1,0},{1,0},{1,-1}}
	
	\newcommand{\segJ}{{0,2},{-2,2}}
	\newcommand{\segK}{{1,1},{0,1},{1,0},{1,0},{1,0}}
	\newcommand{\segL}{{1,-1},{1,-1},{1,0},{1,0}}
	
	\tikzstyle{state}=[circle, fill=gray, inner sep=0pt, minimum size=0.10cm]
	\tikzstyle{important}=[minimum size=0.12cm,black]
	\tikzstyle{reaction}=[line width=0.6mm, densely dotted]
	\tikzstyle{accelerated}=[->,line width=0.7mm]
	
	\tikzstyle{absborder}=[gray, thick]
	
	\newcommand{\absBLcolor}{niceorange!70!orange}
	\newcommand{\absBRcolor}{nicegreen!70!green}
	\newcommand{\absTLcolor}{niceblue!70!blue}
	\newcommand{\absTRcolor}{magenta}
	
	\tikzstyle{abs11color}=[\absBLcolor]
	\tikzstyle{abs12color}=[\absBRcolor]
	\tikzstyle{abs21color}=[\absTLcolor]
	\tikzstyle{abs22color}=[\absTRcolor]
	\tikzstyle{bgOpacity}=[opacity=0.05]

	\drawAbstractState{absborder}{0,0}{7}{7}{0.5}{}{state, important}{abs11}{abs11color, bgOpacity};
	\drawAbstractState{absborder}{0,7}{7}{7}{0.5}{}{state, important}{abs12}{abs12color, bgOpacity};
	\drawAbstractState{absborder}{7,0}{7}{7}{0.5}{}{state, important}{abs21}{abs21color, bgOpacity};
	\drawAbstractState{absborder}{7,7}{7}{7}{0.5}{}{state, important}{abs22}{abs22color, bgOpacity};
	
	\applySegment{state}{reaction, abs11color}{abs11}{abs11a}{\segA};
	\applySegment{state}{reaction, abs11color}{abs11}{abs11b}{\segB};
	\applySegment{state}{reaction, abs11color}{abs11}{abs11c}{\segC};
	\draw[accelerated] (abs11) -- node[pos=0.78,inner sep=0pt,abs11color] {\large$\mathbf{a}$} (abs11a) {};
	\draw[accelerated] (abs11) -- node[pos=0.65,inner sep=0.3pt,abs11color] {\large$\mathbf{b}$} (abs11b) {};
	\draw[accelerated] (abs11) -- node[pos=0.5,inner sep=1.3pt,abs11color] {\large$\mathbf{c}$} (abs11c) {};
	
	\applySegment{state}{reaction, abs21color}{abs21}{abs21a}{\segD};
	\applySegment{state}{reaction, abs21color}{abs21}{abs21b}{\segE};
	\applySegment{state}{reaction, abs21color}{abs21}{abs21c}{\segF};
	\draw[accelerated] (abs21) -- node[pos=0.6,inner sep=0pt,swap,abs21color] {\large$\mathbf{d}$} (abs21a) {};
	\draw[accelerated] (abs21) -- node[pos=0.75,inner sep=1pt,swap,abs21color] {\large$\mathbf{e}$} (abs21b) {};
	\draw[accelerated] (abs21) -- node[pos=0.85,inner sep=1pt,abs21color] {\large$\mathbf{f}$} (abs21c) {};
	
	\applySegment{state}{reaction, abs12color}{abs12}{abs12a}{\segG};
	\applySegment{state}{reaction, abs12color}{abs12}{abs12b}{\segH};
	\applySegment{state}{reaction, abs12color}{abs12}{abs12c}{\segI};
	\draw[accelerated] (abs12) -- node[pos=0.3,inner sep=1pt,swap,abs12color,swap] {\large$\mathbf{g}$} (abs12a) {};
	\draw[accelerated] (abs12) -- node[pos=0.6,inner sep=0pt,abs12color] {\large$\mathbf{h}$} (abs12b) {};
	\draw[accelerated] (abs12) -- node[pos=0.5,inner sep=1pt,swap,abs12color,swap] {$\mathbf{i}$} (abs12c) {};
	
	\applySegment{state}{reaction, abs22color}{abs22}{abs22a}{\segJ};
	\applySegment{state}{reaction, abs22color}{abs22}{abs22b}{\segK};
	\applySegment{state}{reaction, abs22color}{abs22}{abs22c}{\segL};
	\draw[accelerated] (abs22) -- node[pos=0.5,inner sep=0pt,abs22color] {\large$\mathbf{j}$} (abs22a) {};
	\draw[accelerated] (abs22) -- node[pos=0.6,inner sep=1pt,swap,abs22color] {\large$\mathbf{k}$} (abs22b) {};
	\draw[accelerated] (abs22) -- node[pos=0.6,inner sep=1pt,abs22color] {\large$\mathbf{l}$} (abs22c) {};
	
	\begin{scope}[shift=({16.5,0})]
		
		\drawAbstractState{absborder}{0,0}{7}{7}{0.5}{}{state, important}{abs11}{abs11color, bgOpacity};
		\drawAbstractState{absborder}{0,7}{7}{7}{0.5}{}{state, important}{abs12}{abs12color, bgOpacity};
		\drawAbstractState{absborder}{7,0}{7}{7}{0.5}{}{state, important}{abs21}{abs21color, bgOpacity};
		\drawAbstractState{absborder}{7,7}{7}{7}{0.5}{}{state, important}{abs22}{abs22color, bgOpacity};

		\node[state, label={[label distance=-0.05cm,xshift=-0.1cm]-45:{\large $s_{\text{init}}$}}] (step0) at (4,3) {};
		\applySegment{state}{reaction, abs11color}{step0}{step1}{\segC};
		\draw[accelerated] (step0) -- node[pos=0.5,inner sep=1.3pt,abs11color] {\large$\mathbf{c}$} (step1) {};
		\applySegment{state}{reaction, abs21color}{step1}{step2}{\segD};
		\draw[accelerated] (step1) -- node[pos=0.6,inner sep=0pt,swap,abs21color] {\large$\mathbf{d}$} (step2) {};
		\applySegment{state}{reaction, abs12color}{step2}{step3}{\segG};
		\draw[accelerated] (step2) -- node[pos=0.3,inner sep=1pt,swap,abs12color,swap] {\large$\mathbf{g}$} (step3) {};
		\applySegment{state}{reaction, abs12color}{step3}{step4}{\segI};
		\draw[accelerated] (step3) -- node[pos=0.5,inner sep=1pt,swap,abs12color,swap] {$\mathbf{i}$} (step4) {};
		\applySegment{state}{reaction, abs12color}{step4}{step5}{\segH};
		\draw[accelerated] (step4) -- node[pos=0.6,inner sep=0pt,abs12color] {\large$\mathbf{h}$} (step5) {};
		\applySegment{state}{reaction, abs12color}{step5}{step6}{\segI};
		\draw[accelerated] (step5) -- node[pos=0.5,inner sep=1pt,swap,abs12color,swap] {$\mathbf{i}$} (step6) {};
		\applySegment{state}{reaction, abs22color}{step6}{step7}{\segL};
		\draw[accelerated] (step6) -- node[pos=0.6,inner sep=1pt,abs22color] {\large$\mathbf{l}$} (step7) {};
	\end{scope}
\end{tikzpicture}
    \vspace{-0.5em}
    \caption{
        (left) Four neighboring abstract states, drawn as squares. Each abstract state has $\pk{=}3$ segments that start in their respective centers. Each segment is a sequence of reactions drawn as dotted arrows. 
        The difference between the endpoint and the starting point is called a \emph{summary} and is drawn in unbroken black.
        (right)
        A possible segmental simulation obtained by applying the segments $c,d,g,i,h,i$ and $l$ to the initial state $s_{\text{init}}$.
    }
    \label{fig:segmental abstraction}
    \vspace{-1.5em}
\end{figure}

An example for $\pk{=}3$ precomputed segments is depicted in Fig.~\ref{fig:segmental abstraction} (left).
We choose to terminate each segment when it leaves the abstract state. 
Intuitively, at this point,  at least one dimension changes significantly, possibly inducing significantly different rates.

\vspace{-0.5em}
\subsubsection{Assembling the segments.} 
In a \textit{segmental simulation}, instead of sampling a segment for the current concrete state, we sample a segment for the current representative.
Because the sampled segment may start at a different state, we apply the relative effect of the segment to the current concrete state. Note that this is a conceptual difference compared with our previous work~\cite{CAV19} where the segments are applied to abstract states. The importance of this difference is discussed in~\cite[Appendix \ref{appendix:concrete}]{full-version}.
Fig.~\ref{fig:segmental abstraction} (right) illustrates the segmental simulation for the segments on the left. 
The system starts in an initial state, which belongs to the bottom left abstract state. Thus, segment $c$ was randomly chosen among the segments $a,b,c$ belonging to that abstract state. After applying the effect of segment $c$, the system is in the bottom right abstract state and thus we sample from $d,e,f$ and so on. 
Note that applying a segment might not change the current abstract state and might also only do so temporarily (like the application of $l$ and $h$, respectively, in the figure). Once the segment leaves the current state a different set of reactions might be enabled.
Thus, to make sure that we never apply reactions that are not enabled, the population level abstraction has to satisfy some additional constraints.\footnote{We must choose a population abstraction such that applying any of the representative's possible segments to any corresponding concrete state may only change the enabledness of reactions with the last reaction. Similar constraints are needed if we want to avoid transitions to non-neighboring abstract states. For all presented models, the exponential population abstraction with $c{\leq}2$ already has the desired~properties.}

\vspace{-0.5em}
\subsubsection*{Lazy and adaptive computation of segments.}
Instead of precomputing the segments for all abstract states, we generate them on the fly.
When we need to sample the segments of an abstract state $a$ and there are less than $\pk$ segments, then we generate a new segment and store it.
This new segment is the one we would have sampled from the $\pk$ (not yet computed) segments for $a$.
Thus, we enlarge the reservoir of segments lazily, only as we need it.
Since many abstract states might be rarely reached we generate only few segments for them if any at all.
In contrast, we only generate many segments for frequently visited states, which are thus reused many times, improving the efficiency without much overhead. 
Note that segments can be reused already for a single simulation if that simulation visits the same abstract state more than $\pk$ times. However, the real benefit of our approach becomes apparent when we generate many simulations.

\begin{algorithm}[!tb]
    \scriptsize
    \SetKwFunction{isOddNumber}{isOddNumber}
    \SetKwFunction{sampleNewSegment}{sampleNewSegment}
    \SetKwFunction{chooseRandomElementOf}{chooseUniformlyAtRandomFrom}
    \SetKwFunction{abstractState}{abstractState}
    \SetKwFunction{add}{add}
    \SetKwInOut{KwIn}{Inputs}
    \SetKwInOut{KwOut}{Output}

    \KwIn{
    $\crn$ (CRN), $\pk$ (number of segments), $\pc$ (partitioning parameter), \\
    $t_{\text{end}}$ (time horizon), $s_{\text{init}}$ (start state) and $m$ (number of simulations)}
    \KwOut{list of $m$ segmental simulations}

    $simulations := [\ ]$\;
    $memory := \{\}$\tcp*{mapping each abstract state to a list of segments}
    
    \For{$1$ \KwTo $m$}{
        $s := s_{\text{init}}$;\ \ $t := 0$;\ \ $simulation := [(s, t)]$\; 
        \While{$t < t_{\text{end}}$}{
            a := $\abstractState_c(s)$\; \label{alg:main:abstract}
            \eIf{$|memory(a)| < k$}{\label{alg:main:decide}
                $segment := \sampleNewSegment(a.representative)$\tcp*{sample new segment}
                $memory(a).\add(segment)$\tcp*{save it for reuse}\label{alg:main:samplingnew}
            }{
                $segment := \chooseRandomElementOf(memory(a))$\tcp*{reuse old segment}\label{alg:main:precomputed}
            }
            \tcp{apply segment's relative effects}
            $s := s + segment.\Delta_{\text{state}}$;\ \ $t := t + segment.\Delta_{\text{time}}$\;\label{alg:main:acceleration}
            $simulation.\add((s, t))$\;
        }
        $simulations.\add(simulation)$\;
    }
    \KwRet{$simulations$}
    \caption{Lazy Segmental Simulation} 
    \label{alg:main}
\end{algorithm}

\vspace{-1em}
\subsubsection{Algorithm.}
We summarize the approach in the pseudocode of Algorithm \ref{alg:main}.
It is already presented in a way that produces not one but $m$ simulations and computes the segments lazily.
We start with no precomputed segments. 
As we simulate, we always compute the current abstract state $a$ (L.~\ref{alg:main:abstract}) and on L.~\ref{alg:main:decide} decide whether to simulate a new segment (L.~\ref{alg:main:samplingnew}) or uniformly choose from the previously computed ones (L.~\ref{alg:main:precomputed}).

\subsection{Densely Concrete Simulations} \label{ssec:densely}

\subsubsection{Summaries.}
Storing and applying the whole segments can still be memory- and time-intensive. 
Therefore, we replace each segment with a single ``transition'', called a \emph{summary}. 
It captures the overall effect on the state, namely the difference between the end state and the starting state, and the time the sequence of reactions took.
The summaries of segments in Fig.~\ref{fig:segmental abstraction} (left) are depicted as solid black arrows. Algorithm \ref{alg:main} applies the segment's summary in L. \ref{alg:main:acceleration}.

\subsubsection{Assembling summaries.}
\begin{figure}[t]
    \centering
    \includegraphics[width=0.8\textwidth]{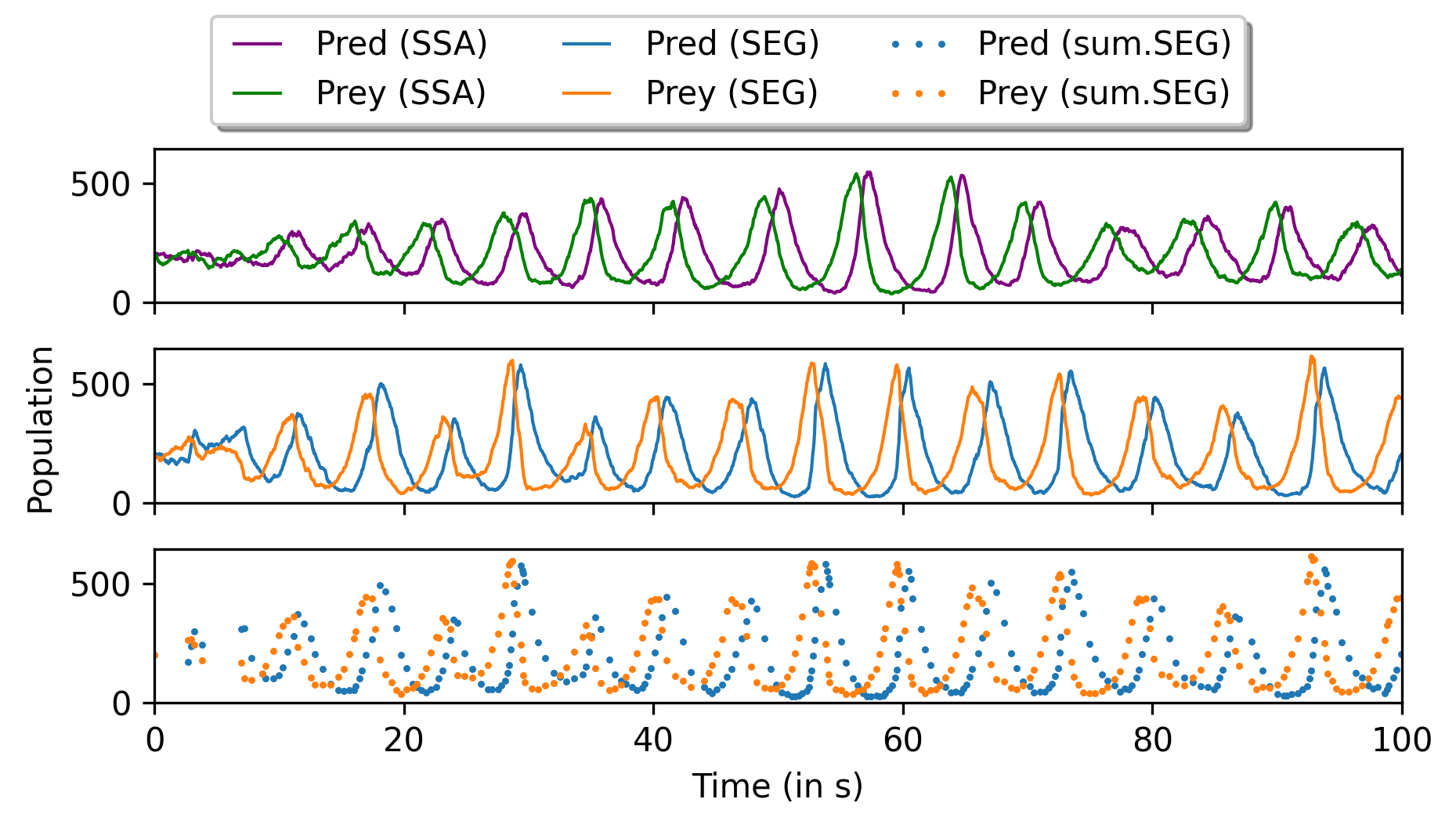}
    \vspace{-1em}
    \caption{Comparison of simulations for the predator-prey system: SSA simulation (top), segmental simulation (middle) and the same segmental simulation as densely concrete simulation where segments are replaced with their summaries (bottom).}
    \vspace{-0.5em}
    \label{fig:pred_prey_examples}
\end{figure}

Instead of segments, we can append their summaries to the simulation runs, see Fig.~\ref{fig:segmental abstraction} (right).
We call the result a \textit{densely concrete simulation}. The effect of this modification can be seen in Fig.~\ref{fig:pred_prey_examples}. 
The top part of the figure shows a typical oscillating run of the predator-prey model that was produced via SSA simulation.
The middle part displays a segmental simulation based on our abstraction that uses segments. It exhibits the expected oscillations with varying magnitudes and is visually indistinguishable from an SSA simulation.
On the bottom, we see the corresponding densely concrete simulation where the segments of the same segmental simulation have been replaced with their summaries.
We still observe the same global behavior but lose the local detail.
More precisely, we only see those concrete states of the middle simulation that are the \emph{seams of the segments} (now displayed as dots).
All the other concrete states are unknown.
However, they are close to the dots since they can only be in the same or neighboring abstract states, meaning the known concrete states are arranged \emph{densely} enough.
Moreover, the distance between the dots corresponds to the lengths of the segments.
Hence the dots are arranged sparsely only if the system entered a very stable abstract state.
Altogether, all changes are reflected faithfully, relative to the level of detail of the abstraction.

\subsection{Introduced inaccuracy}
We summarize the sources of errors our approach introduces.

\vspace{-0.5em}
\subsubsection{(1) Number of segments}
Instead of sampling from all of the possible trajectories, we only sample from $k$ segments and consequently lose some variance.
Thus, if $k$ is too small or if the trajectories are too long, our abstraction might miss important behavior of the original system.
However, it is easy to see that this error vanishes for $\pk \rightarrow \infty$.
It is thus crucial to choose an appropriate value for $k$ and we discuss this choice in Section \ref{sec:evaluation}.
However, note that  sampling from segments instead of reactions cannot produce spurious behavior.
In other words, all trajectories we obtain by sampling segments are  possible trajectories of the original system.
Further, if the segments we sample from are representative enough of the actual distribution of trajectories, we will exhibit the same global behavior.

\vspace{-0.5em}
\subsubsection{(2) Size of the abstract states}
Recall that we do not sample the distribution of segments for the current state but instead sample the distribution for the representative. 
Because the propensities and thus the rates of the reactions are different in the current state and the representative state, this inherently introduces an error.
However, this error is small if we assume that the distribution over segments does not significantly change within the abstract state.
This assumption is reasonable since the propensity functions and thus the rates of the reactions are similar for similar populations and change only slowly; except when the number of molecules is close to zero but there the exponential abstraction provides very fine abstract states.
Further, we can decrease the parameter $\pc$ that determines the interval sizes of the exponential population-level abstraction. 
A discussion of the influence of parameter $\pc$ on the accuracy of our method can be found in Section \ref{sec:evaluation}.

\section{From Segmental Simulations to Abstract Simulations}

In this section, we focus on abstract simulation runs, i.e.~runs over the abstract state space, for several reasons.
Concrete simulation runs, i.e.~runs over the concrete state space, provide a rich piece of information about the system, 
however, already storing a large number of very long simulation runs may be infeasible.
Compared to segmental simulation, densely concrete simulation drops most concrete states by only remembering the seams of the segments; yet the number of concrete states can be large and each one can take non-trivial space if the populations reach large numbers.
Another disadvantage is that, for most of the time points, we only know that the current state is in the same abstract state as the nearest seams or in their neighbors, but no exact concrete state.
In contrast, the abstraction may hide non-interesting details and instead show the big picture.
Altogether, if only population levels are of interest, abstract simulations can be~useful.

\subsection{Segmental Abstraction of CTMC and Abstract Simulation}\label{sec:4.1}

Population-level abstraction of the CTMC might be a lot more explainable than the complete CTMC.
However, while the state space of the population abstraction of a CTMC is simply given by the population levels as the Cartesian product of the intervals over all the species, it is not clear what the nature of the transitions should be.
There are two issues the previous literature faces.
First, what should be the dynamics of an abstract state if each concrete state within behaves a bit differently?
Second, what should be the dynamics of one abstract step between two abstract states when it corresponds to a varying number of concrete steps?
Standard approaches either pick a representative state and copy its dynamics, e.g. the rate of the reaction, or take an over-approximation of all behaviors of all possible members of the class, e.g. take an interval of possible rates.

Here we reuse the segmental simulation and the concepts of Section~\ref{sec:segmental} to formally define a transition function on the abstraction. This gives us the abstraction's semantics and makes it executable.
Moreover, the resulting behavior is close to the original system (in contrast to, e.g.~\cite{CAV19}, we can even preserve the oscillations of the predator-prey models), but at the expense of making the abstract model non-Markovian.
Intuitively, our \emph{segmental abstraction} of the CTMC is given by the abstract state space and the segments, exactly as depicted in Fig.~\ref{fig:segmental abstraction} (left).
Similar to other non-Markovian systems, the further evolution of the system is not given only by its current state, but also by some information about the history of the run so far.
For instance, in the case of semi-Markov processes, it is the time spent in the state so far; in the case of generalized semi-Markov processes, it is the times each event has been already scheduled for.
In the case of the segmental abstraction, it is a vector forming a concrete state.

Formally, the \emph{configuration} of the segmental abstraction is a triple $(a,s,t)$ where $a$ is an abstract state, $s$ one of its concrete states, and $t$ a time.
The probability to move from $(a,s,t)$ to $(a',s',t')$ is then given by the probability to sample a segment with a summary $s'-s$ on states and taking $t'-t$ time.
(Hence we can store the summaries only, as described in Sec.~\ref{ssec:densely}.)
Given a concrete state $s$ to start in, there is a unique probability space over the abstract runs initiated in $(a,s,0)$ obtained by dropping the second component. 

\begin{figure}[t]
    \centering
    \includegraphics[width=0.45\linewidth]{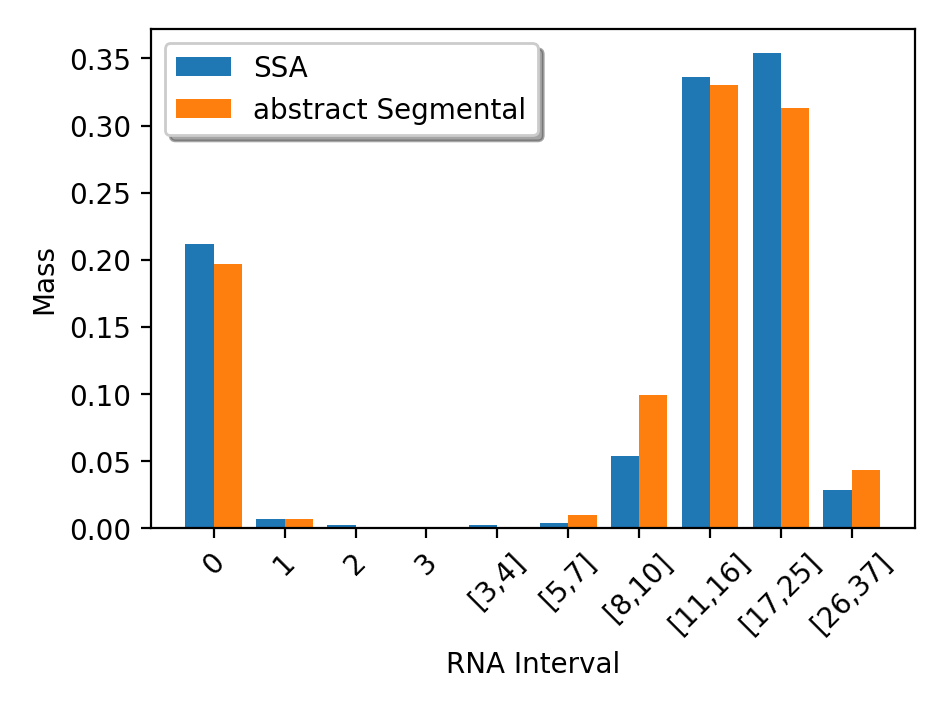}
    \includegraphics[width=0.45\linewidth]{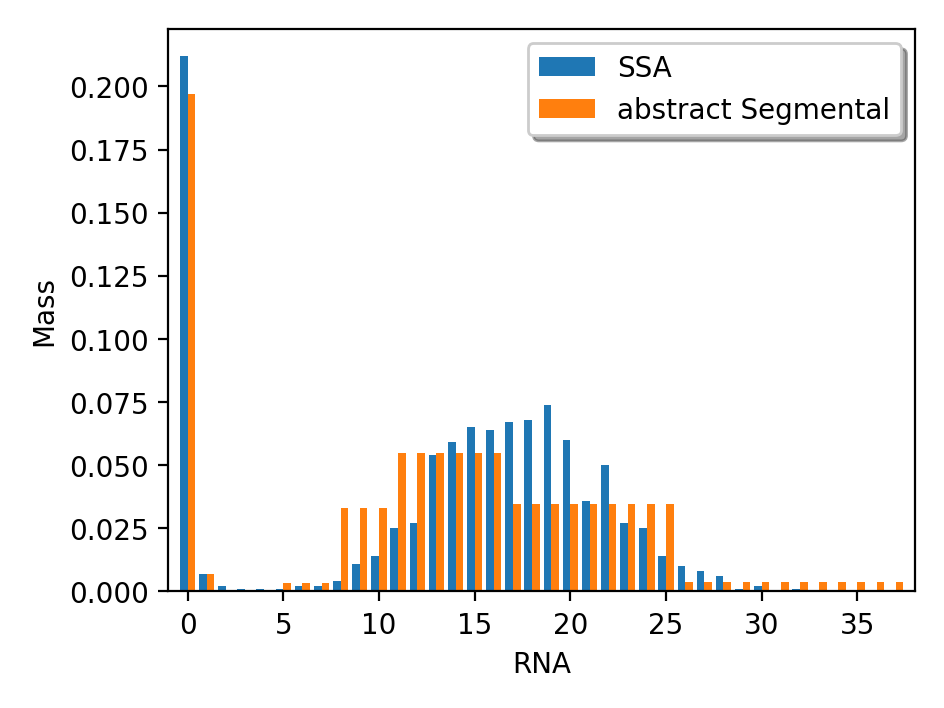}
    \vspace{-1em}
    \caption{RNA distribution in the viral infection model at $t{=}200$s predicted by SSA and abstract segmental simulation with $c{=}1.5$ and $k{=}100$: in the abstract domain (left) and the concrete domain (right) where the segmental simulation's abstract values were concertized using a uniform distribution over the~interval.}
    \vspace{-1em}
    \label{fig:abstract}
\end{figure}

The probability space coincides with the probability space introduced by the segmental simulation (in the variant with summaries of precomputed segments) when the concrete runs are projected by the population-level abstraction to the abstract runs.
However, (i) the space needed to store the abstract simulations is smaller and (ii) transient analysis is well defined for every time point, while its results are still very faithful.
Indeed, Fig.~\ref{fig:abstract} (left) shows an example of the transient analysis (at a given time point $t$) obtained by (i) the states reached by real simulation runs and clustered according to the population-level abstraction, and (ii) abstract states reached by abstract segmental simulation runs.
Given the granularity of the abstraction, the results are very close.

\subsection{From abstract simulations back to concrete predictions}\label{sec:4.2}

Further, one can map the abstract states to sets of concrete states.
Consequently, the results of the abstract transient analysis can be mapped to a distribution over concrete states, whenever we assume a distribution over the concrete states corresponding to one abstract state.
For instance, taking uniform distribution as a baseline, we obtain a concrete transient analysis from the abstract one, see Fig.~\ref{fig:abstract} (right), which already shows a close resemblance.

\section{Experimental evaluation}\label{sec:evaluation}

We evaluate the densely concrete version of segmental simulation and consider the following three research questions:

\begin{itemize}
    \item[Q1] What is the accuracy of segmental simulation?
    \item[Q2] What are the trade-offs between accuracy and performance?
    \item[Q3] Are the achieved trade-offs competitive with alternative approaches?
\end{itemize}

\subsection*{Experimental setting and accuracy measurement}

\paragraph{Benchmark selection.} We use the following models from the literature: (1) Viral Infection (VI)~\cite{srivastava2002stochastic}, (2) Repressilator (RP)~\cite{hepp2015adaptive}, (3) Toggle Switch (TS)~\cite{hepp2015adaptive} and (4) Predator-Prey (PP, a.k.a. Lotka–Volterra)~\cite{gillespie1977exact}. 
The formal definition for each of these models can be found in~\cite[Appendix~\ref{appendix:models}]{full-version}.
Although the underlying CRNs are quite small (up to 6 species and 15 reactions), their analysis is very challenging due to stochasticity, multi-scale species populations and stiffness.
Therefore, the models are commonly used to evaluate advanced numerical as well as simulation methods. 

\vspace{-0.2em}
\paragraph{Implementation and HW configuration.} Our approach is implemented as modification of \href{https://sequaia.model.in.tum.de/}{SeQuaiA}, a Java-based tool for semi-quantitative analysis of CRNs \cite{CAV20}.
All experiments run on a 1.80 GHz Lenovo ThinkPad T580 with 8GB of RAM. 
To report speedups, we use our own competitive implementation of the SSA method as a baseline. Depending on the model, it achieves between $1{\times}10^6$ and $7{\times}10^6$ reactions per second. We refer to the SeQuaiA repository\footnote{\href{https://sequaia.model.in.tum.de}{https://sequaia.model.in.tum.de} (SeQuaiA)} for all active development 
and provide an \href{https://doi.org/10.5281/zenodo.6658924}{artefact}\footnote{\href{https://doi.org/10.5281/zenodo.6658924}{https://doi.org/10.5281/zenodo.6658924} (artifact)} to reproduce the experimental results.

\vspace{-0.2em}
\paragraph{Assessing accuracy.} 
To measure the accuracy, we compare transient distributions for one species at a time. 
For this, we approximate the implied transient distribution of our approach and of SSA by running a large number of simulations. We used 1000 simulations for the models RP and TS as their simulations take longer, and 10000 for PP and VI. The resulting histograms for the studied species are then normalized to approximate the transient distribution of that species. To quantify the error, we compare the means of both distributions and report the earth-mover-distance (EMD) between them. 
Because EMD values are difficult to interpret without context, we additionally report the EMD between two different transient distributions that were computed with SSA. Intuitively, even if segmental simulation was as accurate as SSA, we would expect to see a EMD similar to the EMD of this "control SSA". Additionally, Fig.~\ref{fig:variance} in \cite[Appendix \ref{appendix:quantitative}]{full-version} compares the variance.

\vspace{-0.2em}
\subsection*{Q1 What is the accuracy of segmental simulation?}

In this section, we evaluate the accuracy of the segmental simulation scheme and the effects of parameters $c$ (size of the abstraction) and $k$ (the number of stored summaries) in a quantitative manner. For a more qualitative evaluation see Fig.~\ref{fig:pred_prey_examples} and~\cite[Appendix~\ref{appendix:qualitative}]{full-version} where you find exemplary simulations and trajectories for both segmental simulation and SSA.
We consider three population abstractions given by $c\in \{2,1.5,1.3\}$ (recall that $c{=}2$ is the most coarse abstraction as explained in Section 2) and three values of $k$, namely  $k\in\{10,100,1000\}$.

\begin{figure}[tb]
    \centering
    \begin{minipage}[t]{.47\textwidth}
        \includegraphics[width=\linewidth]{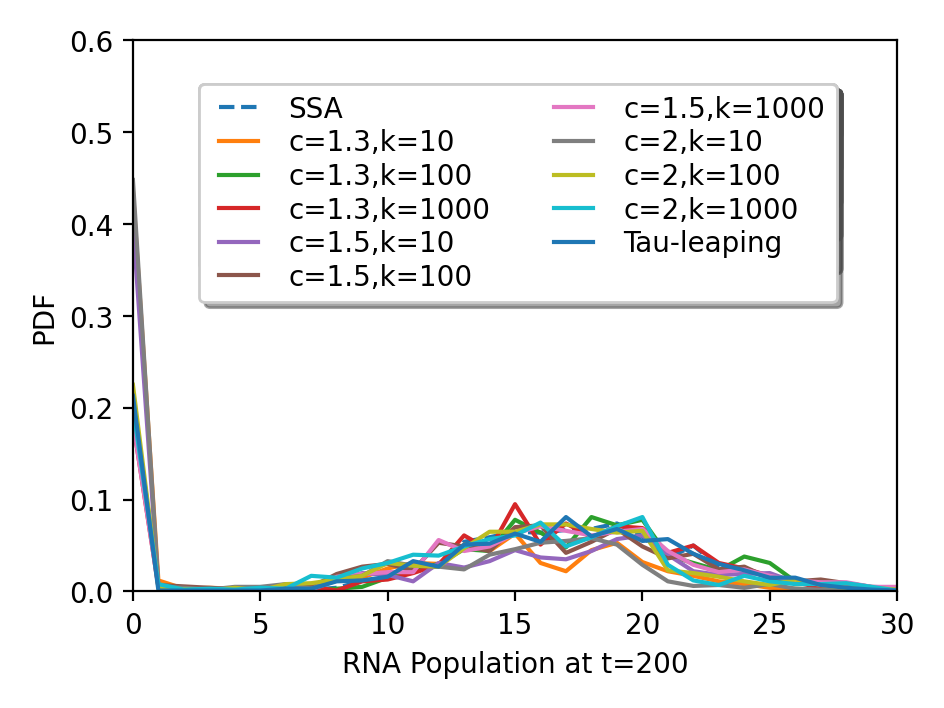}
    \end{minipage}
    \begin{minipage}[t]{.51\textwidth}
        \includegraphics[width=\linewidth]{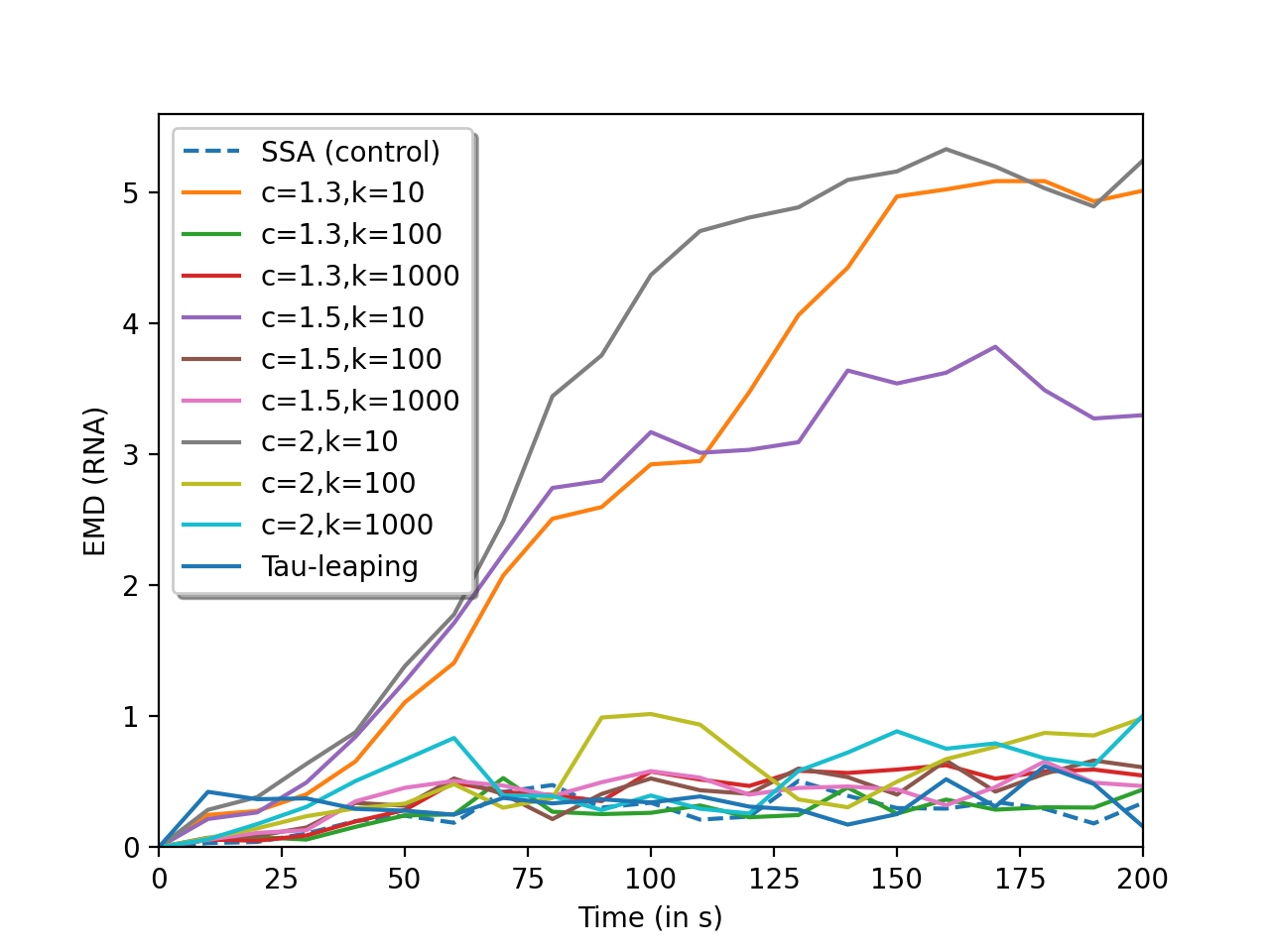}
    \end{minipage}
    \vspace{-1.5em}
    \caption{Accuracy on the viral infection model using different abstractions.}
    \label{fig:viral}
    \vspace{-1em}
\end{figure}

We start with the VI model where one is typically interested in the distribution of the RNA population at a given time $t$. Fig.~\ref{fig:viral} (left) shows the distributions at $t{=}200$ obtained by SSA simulation and by segmental simulations with different values of the parameters $c$ and $k$. We observe that all distributions show the expected bi-modality~\cite{goutsias2005quasiequilibrium}. If a less precise abstraction is used ($c{=}2$ and/or $k{=}10$), the probability that RNA dies out is significantly higher than the reference value. In Fig.~\ref{fig:viral} (right), we evaluate how the EMD (for the RNA) changes in time for particular settings. The results clearly confirm that $k{=}10$ leads to significant inaccuracy. For all other settings, the EMD is very close to the SSA control demonstrating the very high accuracy of our approach. The only notable exceptions are the variants with $c{=}2$, where the EMD fluctuates. We also observe that increasing $k$ from 100 to 1000 does not bring any considerable improvement.

\begin{figure}[tb]
    \centering
     \includegraphics[width=0.49\linewidth]{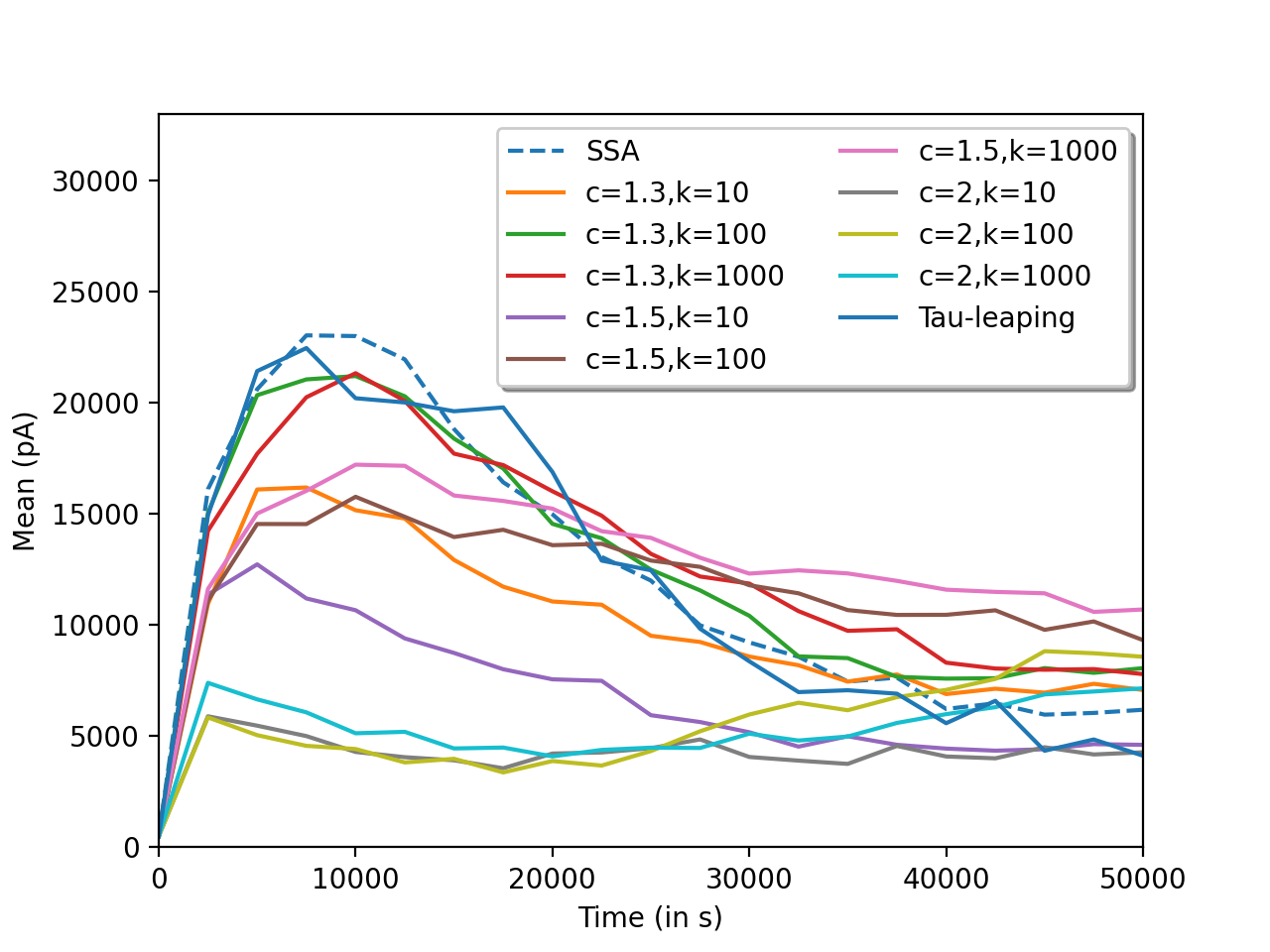}
     \includegraphics[width=0.49\linewidth]{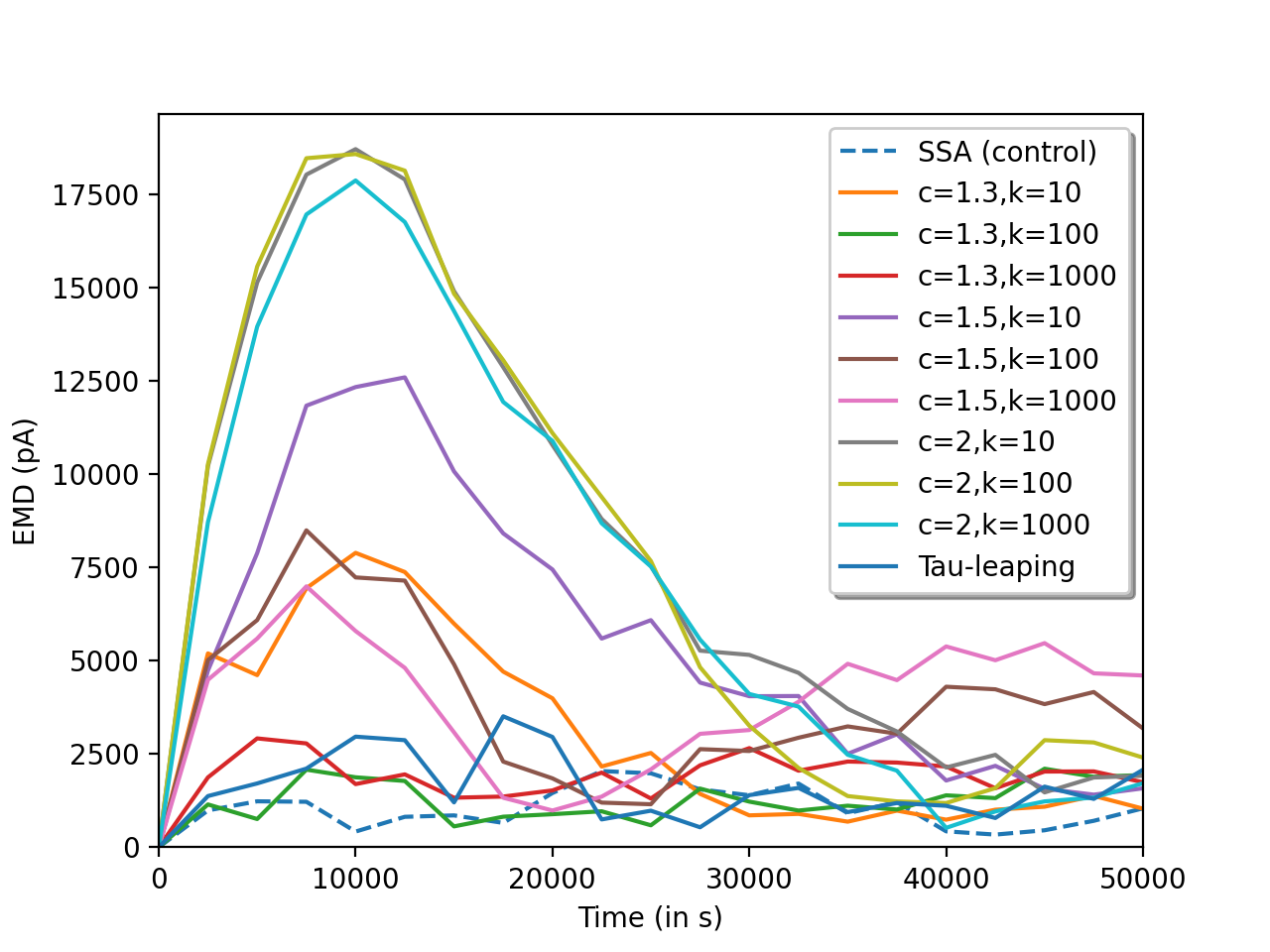}
     \vspace{-1em}
        \caption{Accuracy on the repressilator model using different abstractions.}
            \label{fig:Rep}
            \vspace{-1em}
\end{figure}

Different trends can be observed for the RP model. Fig.~\ref{fig:Rep} shows how the mean value (left) and the EMD (right) of the species pA change over time. We observe that the partitioning of the populations plays a more important role here, i.e., the coarse abstraction ($c{=}2$) induces a notable inaccuracy. The fact that the accuracy is less sensitive with respect to the low values of $k$ is a result of the very regular dynamics of the model where the populations of the proteins pA and pB oscillate and slowly decrease. Similar trends (not presented here) are observed also for the TS model.

\begin{figure}[tb]
    \centering
     \includegraphics[width=0.50\linewidth]{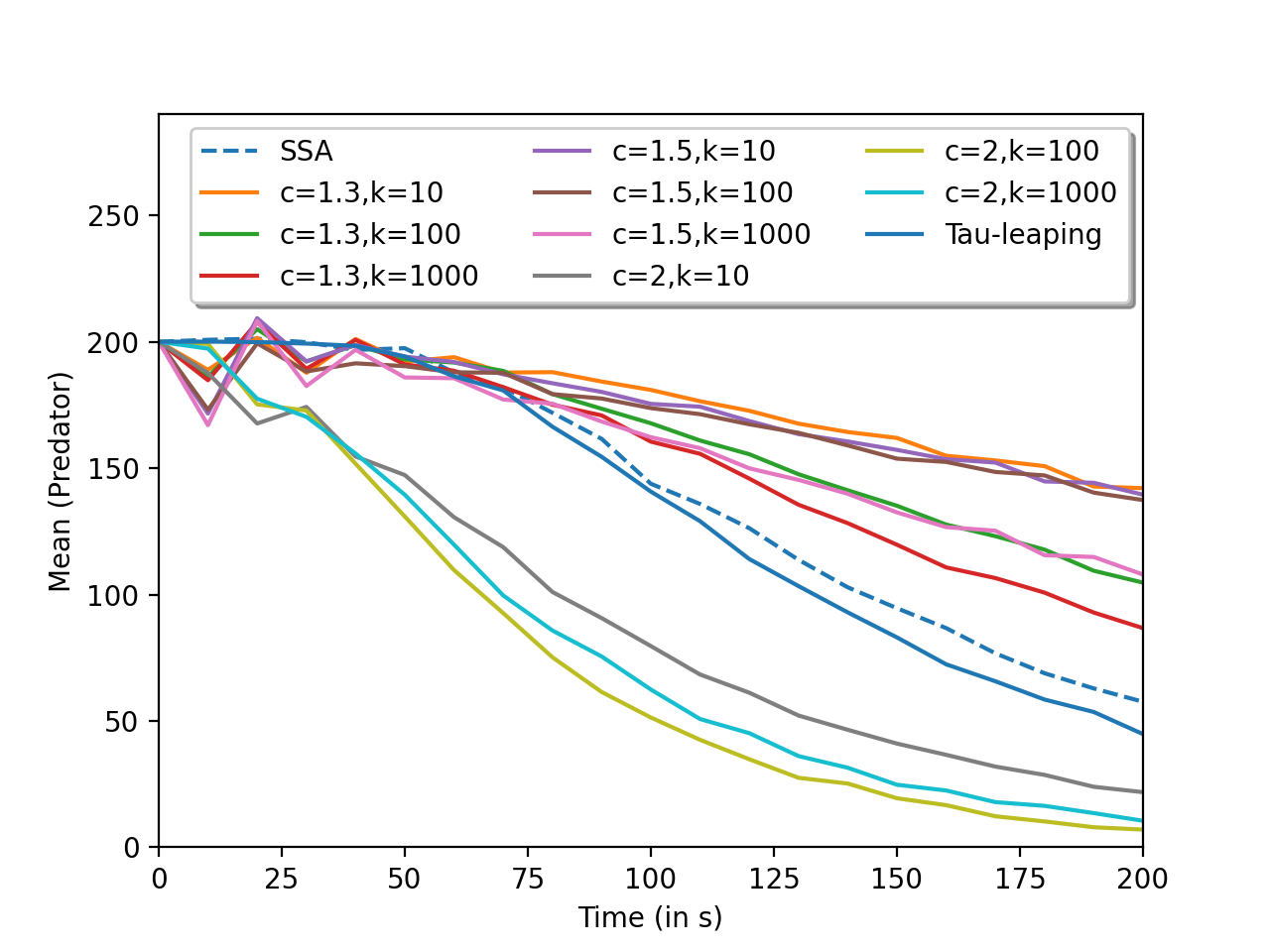}
     \includegraphics[width=0.47\linewidth]{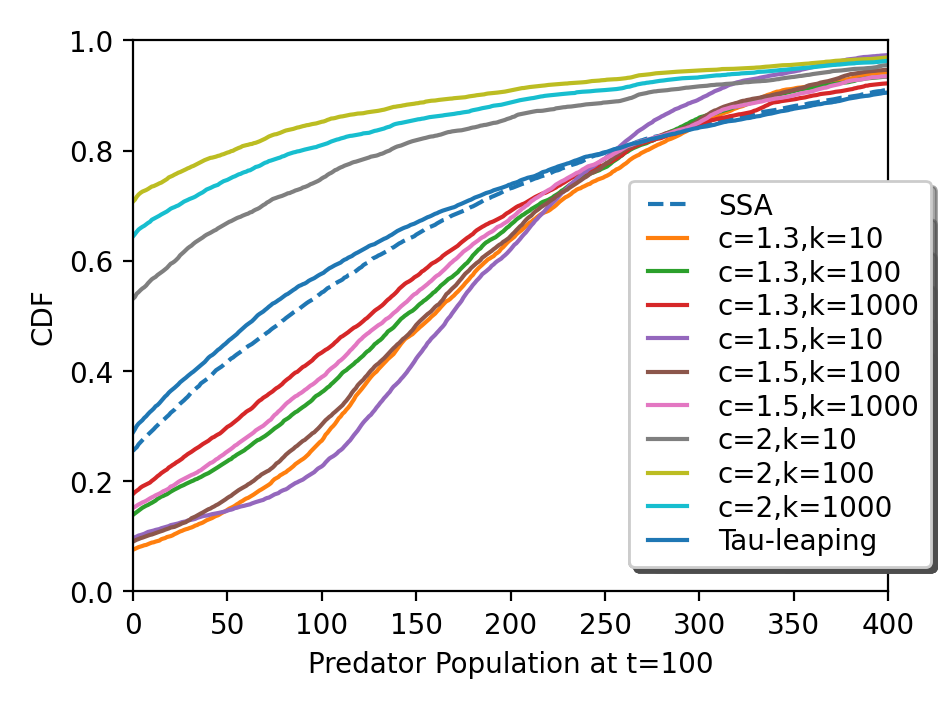}
     \vspace{-1em}
        \caption{Accuracy on the predator-prey model using different abstractions.}
            \label{fig:Pred}
            \vspace{-1.2em}
\end{figure}

Finally, we consider the PP model. Although very simple, it is notoriously difficult for abstraction-based approaches since they  struggle to preserve the oscillation and the die-out time. Recall Fig.~\ref{fig:pred_prey_examples} of Section \ref{sec:segmental} where we clearly observe the expected oscillation with the correct frequency in a segmental simulation for $c{=}2$ and $k{=}100$. Fig.~\ref{fig:Pred} (left) shows how the mean population of the predators changes in time. We observe that the less precise abstractions do not accurately preserve the rate at which the mean population decreases. On the other hand, the most precise setting is close to the SSA reference and control curve.  Fig.~\ref{fig:Pred} (right) shows the cumulative predator distributions at $t{=}100$ demonstrating how the simulations using less precise abstractions deviate from the reference solution. 

\vspace{-0.5em}
\subsection*{Q2 What are the trade-offs between accuracy and performance?}

\begin{figure}[t]
    \centering
    \includegraphics[width=0.49\linewidth]{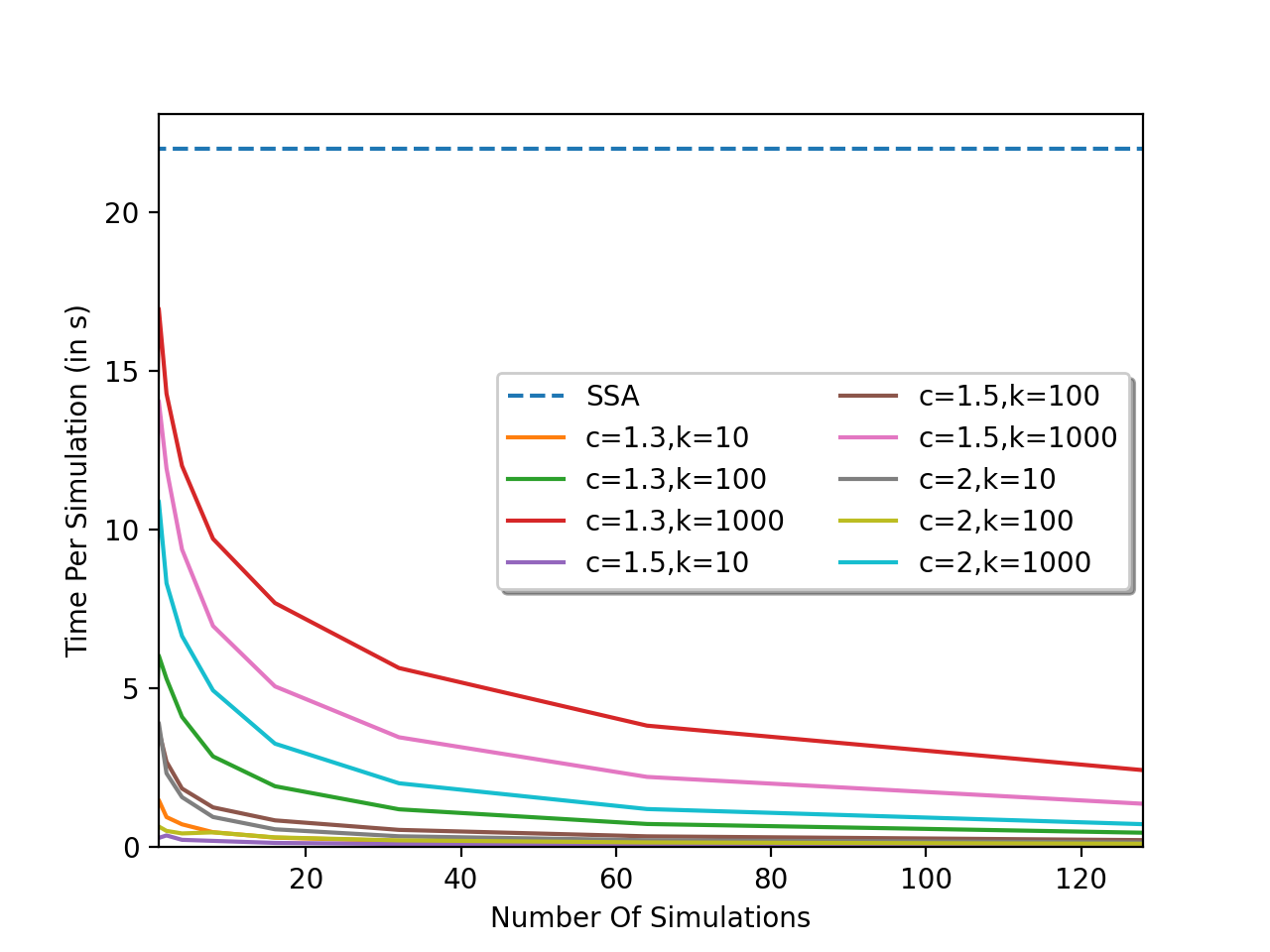}
    \includegraphics[width=0.49\linewidth]{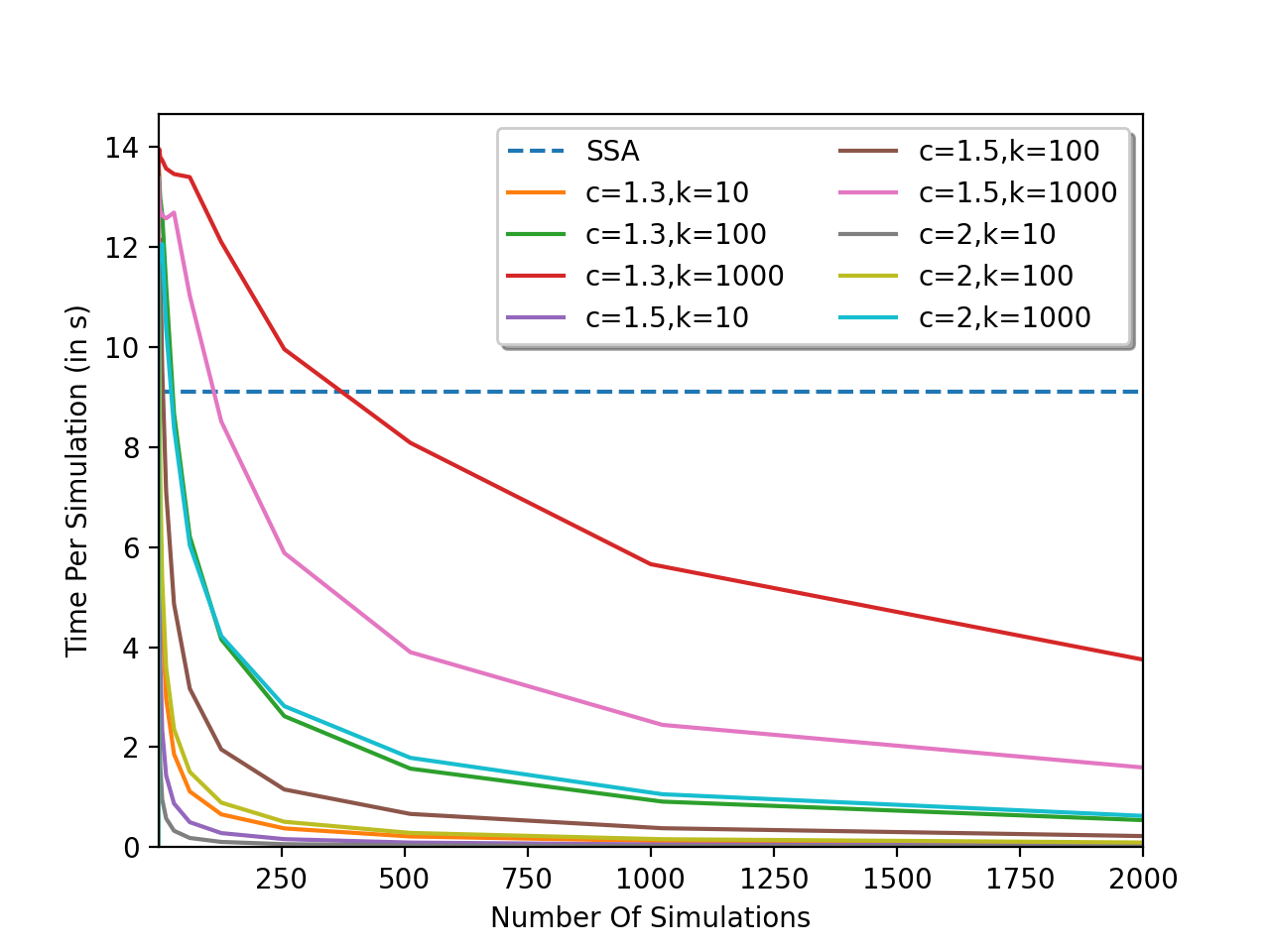}
    \vspace{-1em}
    \caption{Increasing speed of lazy segmental simulation for the toggle switch model (left) and repressilator model (right).}
    \label{fig:speedup}
    \vspace{-1em}
\end{figure}

Recall that our approach is based on re-using the segments generated in the previous simulation runs. Fig.~\ref{fig:speedup} shows how the average time per simulation decreases for a growing number of simulations. For some models (like TS) segmental simulation is always faster than SSA because it can reuse segments already during the first simulation. For other models (like RP), segmental simulation becomes faster after a number of simulations that depends on the precision of the abstraction.
Table~\ref{tab:time} shows the speedup factor we observe when running 10,000 segmental simulations instead of SSA simulations. Table~\ref{tab:parts} in \cite[Appendix \ref{appendix:quantitative}]{full-version} reports the average number of reactions per SSA simulation and the average number of applied summaries per segmental simulation. 
Comparing these numbers highlights the source of the speedup: instead of sampling and applying many reactions, segmental simulation does the same for fewer summaries. This also gives an estimate for the speedup factor that can be reached once only precomputed summaries are used. Note that the speedup factors we report are smaller because we include the time needed for computing the summaries. 

The PP model includes only two species and it is not stiff. Therefore, already the SSA simulation is quite fast. 
Our approach still achieves a stable speedup around 70x that drops to a factor of 12x for the most precise abstraction ($c{=}1.3$ and $k{=}1000$) providing accuracy close to the control SSA. 

For the VI model, we observe a slowdown of the segmental simulation when improving the abstraction, namely, for $k{=}1000$. Recall, we reported very good accuracy already for $c{=}2$ and $k{=}100$ which gives us a speedup factor of 100x. 
For an accuracy that is close to the control SSA, we archive a speedup~factor~of~50x.

The TS model exhibits regular oscillation, where a typical run repeatedly visits the same abstract states. This is very beneficial for our approach as we can very efficiently reuse segments.  
We observe a significant slowdown for $c{=}1.3$ and $k{=}1000$, since reusing segments is much less effective. The RP model has similar characteristics, but we observe an even more significant slowdown when the abstraction is refined. As we discussed in the previous section, very good accuracy for these models is achieved already for $c{=}1.5$ and $k{=}100$ which give us speedup factors 350x and 140x for the TS and RP model, respectively. 

\begin{table}[t]
    \centering
    \begin{tabular}{c||c||r|r|r||r|r|r||r|r|r|}
    \multirow{2}{*}{Model}  &     \multirow{2}{*}{SSA}        & \multicolumn{3}{c||}{SEG $k=10$} &  \multicolumn{3}{c||}{SEG $k=100$ } &  \multicolumn{3}{c|}{SEG $k=1000$ }\\ \cline{3-11}
    &  & \multicolumn{1}{c|}{$c=2$} & \multicolumn{1}{c|}{$c=1.5$} & \multicolumn{1}{c||}{$c=1.3$} & \multicolumn{1}{c|}{$c=2$} & \multicolumn{1}{c|}{$c=1.5$} & \multicolumn{1}{c||}{$c=1.3$}  & \multicolumn{1}{c|}{$c=2$} & \multicolumn{1}{c|}{$c=1.5$} & \multicolumn{1}{c|}{$c=1.3$} \\ \hline \hline
    PP & 0.014s & 70x & 70x & 70x & 70x & 70x & 23x & 28x & 23x & 12x \\ \hline
    VI & 0.88s & 730x & 380x & 180x & 100x & 48x & 17x & 8.6x & 4.8x & 2.9x \\ \hline
    TS & 22s & 360x & 360x & 340x & 390x & 350x & 280x & 250x & 190x & 110x \\ \hline
    RP & 9.1s & 760x & 540x & 320x & 300x & 140x & 62x & 54x & 21x & 7.4x \\ \hline

    \end{tabular}
    \vspace{1em}
    \caption{Average run-time of one SSA simulation and the speedup factor of segmental simulation when computing 10,000 simulations with different abstraction parameters.}
    \label{tab:time}
      \vspace{-1.2em}
\end{table}

\begin{table}[t]
    \centering
    %\begin{tabular}{c||c|c|c||c|c|c||c|c|c|}
    \begin{tabular}{c||r|r|r||r|r|r||r|r|r|}
     \multirow{2}{*}{Model} & \multicolumn{3}{c||}{SEG $k=10$} & \multicolumn{3}{c||}{SEG $k=100$ } &  \multicolumn{3}{c|}{SEG $k=1000$ }\\ \cline{2-10}
     & \multicolumn{1}{c|}{$c=2$} & \multicolumn{1}{c|}{$c=1.5$} & \multicolumn{1}{c||}{$c=1.3$} & \multicolumn{1}{c|}{$c=2$} & \multicolumn{1}{c|}{$c=1.5$} & \multicolumn{1}{c||}{$c=1.3$}  & \multicolumn{1}{c|}{$c=2$} & \multicolumn{1}{c|}{$c=1.5$} & \multicolumn{1}{c|}{$c=1.3$} \\ \hline \hline 
PP & 163 & 398 & 832 & 170 & 391 & 888 & 170 & 397 & 885\\ \hline
VI & 1,022 & 3,669 & 1,0337 & 1,269 & 3,797 & 11,524 & 1,353 & 4,018 & 11,218\\ \hline
TS & 5,072 & 13,167 & 38,827 & 9,248 & 25,733 & 65,259 & 10,388 & 29,424 & 74,950\\ \hline
RP & 12,921 & 42,241 & 124,280 & 18,014 & 56,312 & 155,394 & 21,460 & 64,385 & 146,126\\ \hline
    \end{tabular}
    \vspace{1em}
    \caption{Number of visited abstract states after 10,000 segmental simulations for different abstraction parameters.}
    \label{tab:abs_states}
      \vspace{-1.8em}
\end{table}

\begin{table}[t]
    \centering
    \begin{tabular}{c||r|r|r||r|r|r||r|r|r|}
    \multirow{2}{*}{Model} & \multicolumn{3}{c||}{SEG $k=10$} &  \multicolumn{3}{c||}{SEG $k=100$ } &  \multicolumn{3}{c|}{SEG $k=1000$ }\\ \cline{2-10}
    & \multicolumn{1}{c|}{$c=2$} & \multicolumn{1}{c|}{$c=1.5$} & \multicolumn{1}{c||}{$c=1.3$} & \multicolumn{1}{c|}{$c=2$} & \multicolumn{1}{c|}{$c=1.5$} & \multicolumn{1}{c||}{$c=1.3$}  & \multicolumn{1}{c|}{$c=2$} & \multicolumn{1}{c|}{$c=1.5$} & \multicolumn{1}{c|}{$c=1.3$} \\ \hline \hline 
PP & 25 KB & 61 KB & 130 KB & 250 KB & 570 KB & 1.3 MB & 2.2 MB & 4.8 MB & 11 MB\\ \hline
VI & 210 KB & 730 KB & 2.0 MB & 1.8 MB & 4.8 MB & 13 MB & 11 MB & 25 MB & 53 MB\\ \hline
TS & 1.2 MB & 3.0 MB & 8.7 MB & 15 MB & 37 MB & 85 MB & 100 MB & 250 MB & 550 MB\\ \hline
RP & 3.8 MB & 12 MB & 34 MB & 43 MB & 120 MB & 300 MB & 310 MB & 760 MB & 1.0 GB\\ \hline
    \end{tabular}
    \vspace{0.5em}
    \caption{Size of segmental abstraction after 10,000 simulations for different parameters.}
    \label{tab:mem_size}
      \vspace{-2.2em}
\end{table}

In general, segmental simulation is never slower than SSA by more than the constant factor that is the result of saving and loading segments and it eventually becomes faster if we compute enough simulations.
We pay for the inevitable speedup with increased memory consumption. Table~\ref{tab:abs_states} shows how the number of visited abstract states after 10,000 segmental simulations. The number of saved summaries \cite[Appendix \ref{appendix:quantitative}, Table~\ref{tab:summaries}]{full-version} is approximately the number of visited abstract states times $\pk$. The memory consumed by one summary is an integer vector and a floating-point number that describe the effect on state and time, respectively.
The size of the abstraction is shown in Table~\ref{tab:mem_size}.

We observe a trade-off between accuracy and performance: smaller abstract states results in segments with fewer steps and thus slower simulations and for more precise abstractions we need to save more summaries.
Further, we can only reuse simulations if they visit the same abstract states. This implies that the presented approach does not scale favorably with the dimension of the studied system. 
To handle many species, one can use the available memory only for the most important abstract states, e.g.~the most visited ones, and simulate normally in all other regions.
But there is an inherent trade-off between memory consumption and simulation speed as we can only reuse segments if we save them.

\vspace{-0.5em}
\subsection*{Q3: Comparison with alternative approaches}

\paragraph{Comparison with $\tau$-leaping.} 

We first compare the performance and accuracy of our approach with the $\tau$-leaping method implemented in StochPy~\cite{maarleveld2013stochpy}, a widely used stochastic modeling and simulation package. $\tau$-leaping achieves very good accuracy on the considered models. Quantitatively, it is very close to the SSA control runs and typically provides slightly better results than our best setting (compare Figs. \ref{fig:viral} to \ref{fig:Pred}). On the other hand, we typically observe only a moderate speedup (around one order of magnitude) with respect to the SSA. Note that
a direct comparison with our run-times is unfair as it is known  that StochPy uses an inefficient  python-based random number generator that can significantly slow down the simulation. For example, a single SSA simulation of the RP model takes in StochPy 1000s and $\tau$-leaping achieves a 16-times speedup while our SSA baseline takes around 9 seconds and the segmental simulation achieves the speedup of a factor over 140  (to our baseline) with only a small drop in the accuracy. On the other hand, for the PP model, $\tau$-leaping provides only a negligible speedup (below factor 2). Recall we observed a speedup factor between 12x and 70x depending on the required precision.

\paragraph{Comparison with advanced simulation methods.}

A fair comparison with slow-scale stochastic simulations~\cite{rao2003stochastic,cao2005slow,goutsias2005quasiequilibrium} is problematic since, to our best knowledge, there is no available implementation. 
Therefore, we focus on the comparison with the results presented in~\cite{hepp2015adaptive} representing the state-of-the-art hybrid simulation method. In Table~\ref{tab:comp}, we compare the run-times of the  adaptive hybrid simulation (100,000 runs) on the repressilator and toggle switch models reported in~\cite{hepp2015adaptive} (Fig.~1) with our approach. We report the runtimes only for the setting with $c{=}1.5$ and $k{=}100$ which already leads to very good accuracy for these models. The table also shows runtimes for the baseline SSA and the achieved speedup factor to make the comparison between the different hardware configurations fair. We observe that a significant computational gain (over two orders of magnitude) is achieved by our approach. 

\begin{table}[t]
    \centering
    \begin{tabular}{c||c|c|c||c|c|c|}
     \multirow{2}{*}{Model}& \multicolumn{3}{c||}{results presented in~\cite{hepp2015adaptive}}  & \multicolumn{3}{c|}{Our results} \\ \cline{2-7}
      & SSA & Adaptive hybrid & Speedup   & SSA & SEG($c{=}1.5$,$k{=}100$) & Speedup \\ \hline \hline
     RP & 232 hours & 3 hours & 77x & 252 hours & 1.8 hours & 140x \\ \hline
     TS & 47 days & 1.1 days & 43x & 25 days & 0.07 days & 350x \\ \hline
    \end{tabular}
    \vspace{1em}
    \caption{Runtime comparison with~\cite{hepp2015adaptive} for 100,000 simulations.}
    \vspace{-2em}
    \label{tab:comp}
\end{table}

\vspace{-0.3em}
\paragraph{Comparison with the deep learning approaches.} Finally, we compare with the approach of~\cite{cairoli2021abstraction} where a neural network is trained to provide a fast and accurate generator of simulations in the original CRN. To this end, we use a simpler variant of the toggle switch model considered in~\cite{cairoli2021abstraction}. If we run more than 1000 simulations, a single simulation takes on average 0.0004s which is comparable with the neural-based generator (the authors report 0.0008s per simulation). Regarding the accuracy, we achieve comparable values of the EMD (note the EMD is scaled in \cite{cairoli2021abstraction}). The key benefit of our approach, however, lies in the fact that it does not require the computationally very demanding training phase. 

\vspace{-0.5em}
\section{Conclusion and Future Directions}
We have proposed a novel simulation scheme enabling us to efficiently generate a large number of simulation runs for complex CRNs. It is based on reusing segments of the runs computed over abstract states but applied to concrete states.
Already our initial experiments demonstrate 
that the simulation scheme preserves key dynamical and quantitative properties while providing a significant computational gain over the existing approaches. 
On the conceptual level, we define an executable abstraction of the CTMC, preserving the dynamics very faithfully.
In particular, we have the machinery to generate abstract simulation runs, which take less space than the concrete ones, yet provide high precision on the level of detail given by the population levels defined by the user.

In future work, we want to investigate the error with the goal of giving formal error bounds. Further, we propose an adaptive version of the abstraction where the population abstraction and the number of precomputed segments are refined or learned.
Alternatively, instead of memorizing a discrete distribution over precomputed segments, we can generalize to unobserved behavior by learning some continuous distribution.

Segmental simulation via abstract states can be understood as a general framework for accelerating the simulation of population models. As such, it can be combined with any method that predicts the evolution of such models. In particular, it can naturally leverage an adaptive multi-scale approach where different simulation techniques are used in different regions of the state-space. 

\bibliographystyle{splncs04}
\bibliography{bib,bib2,bib3,new_bib}

\newpage

\section*{Appendix}

\appendix

\appendix

\section{Qualitative comparison of segmental simulation} \label{appendix:qualitative}

\begin{figure}[!h]
    \centering
    \includegraphics[width=0.45\linewidth]{fig/PP/visual_comparison_c2_k100.png}
    \includegraphics[width=0.45\linewidth]{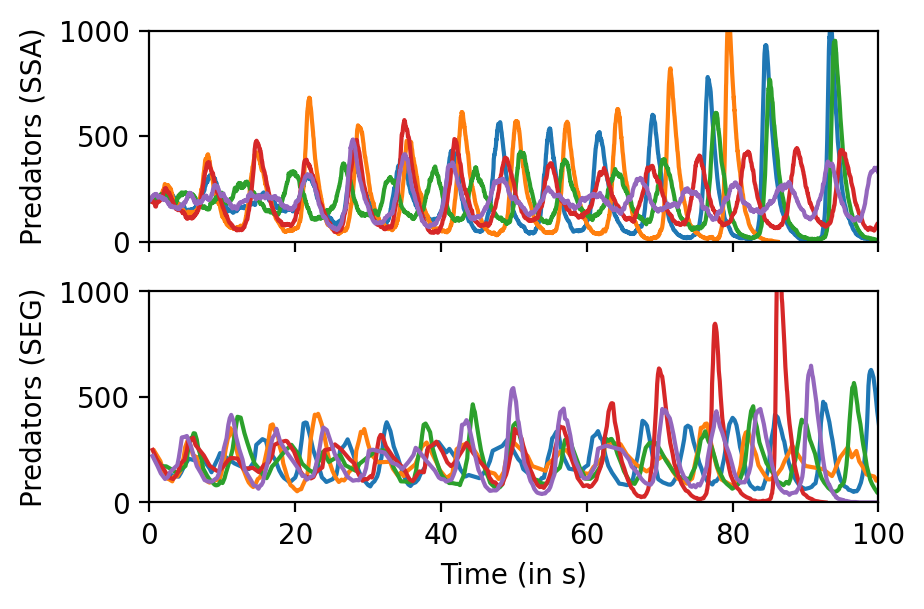}
    \caption{Visual comparison of simulations for the predator-prey model: (left) the top SSA simulation with a segmental simulation below, once with segments and once summaries; (right) five trajectories of Predator species for SSA and segmental simulation with $\pc{=}2$ and $\pk{=}100$.}
    \label{fig:pred_pray_sims}
\end{figure}

\begin{figure}[!h]
    \centering
    \includegraphics[width=0.45\linewidth]{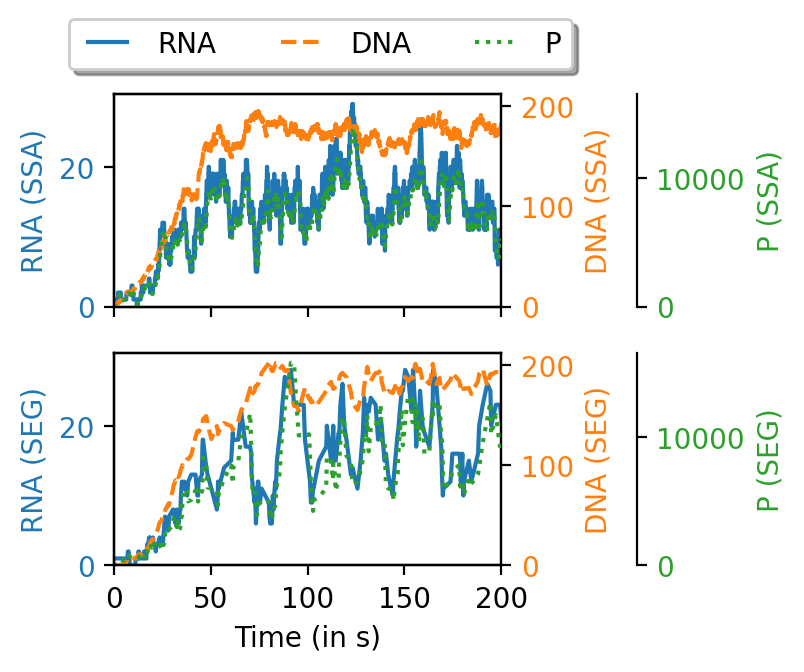}
    \includegraphics[width=0.45\linewidth]{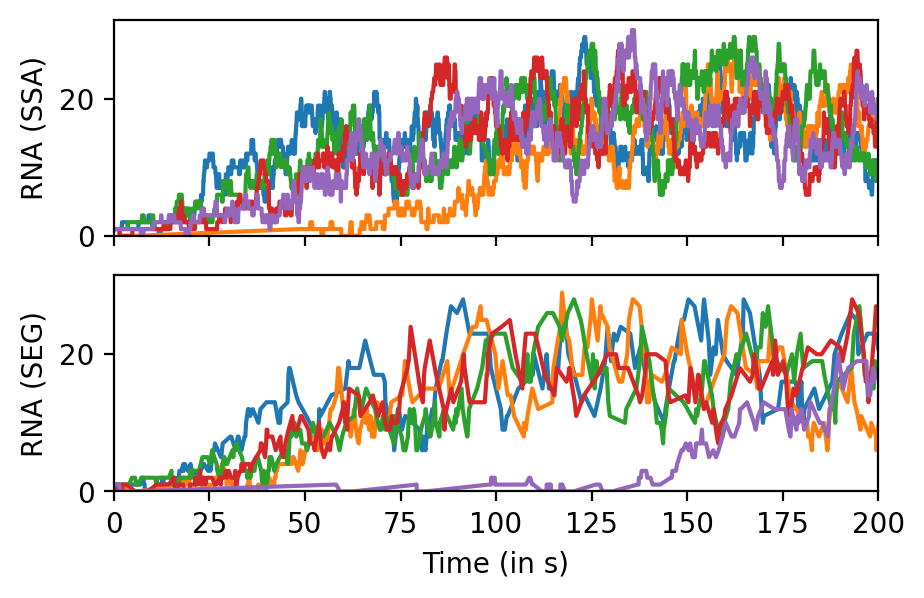}
    \caption{Visual comparison of simulations for the viral infection model (left) two simulations with multiple species (right) five trajectories of RNA species for SSA and segmental simulation with $\pc{=}1.5$ and $\pk{=}100$.}
    \label{fig:viral_sims}
\end{figure}

\clearpage

\begin{figure}[!h]
    \centering
    \includegraphics[width=0.45\linewidth]{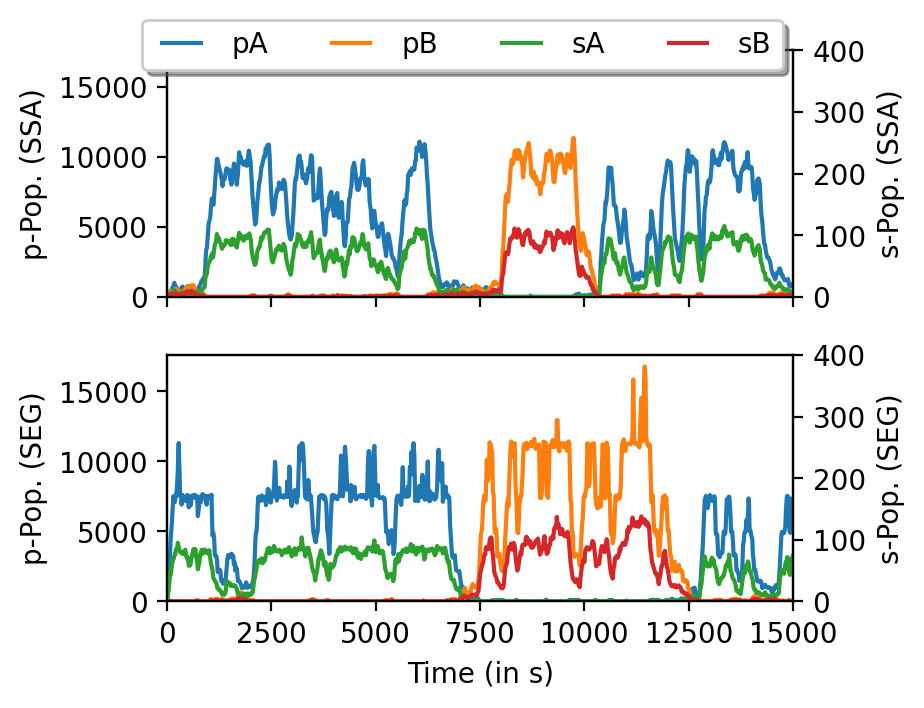}
    \includegraphics[width=0.45\linewidth]{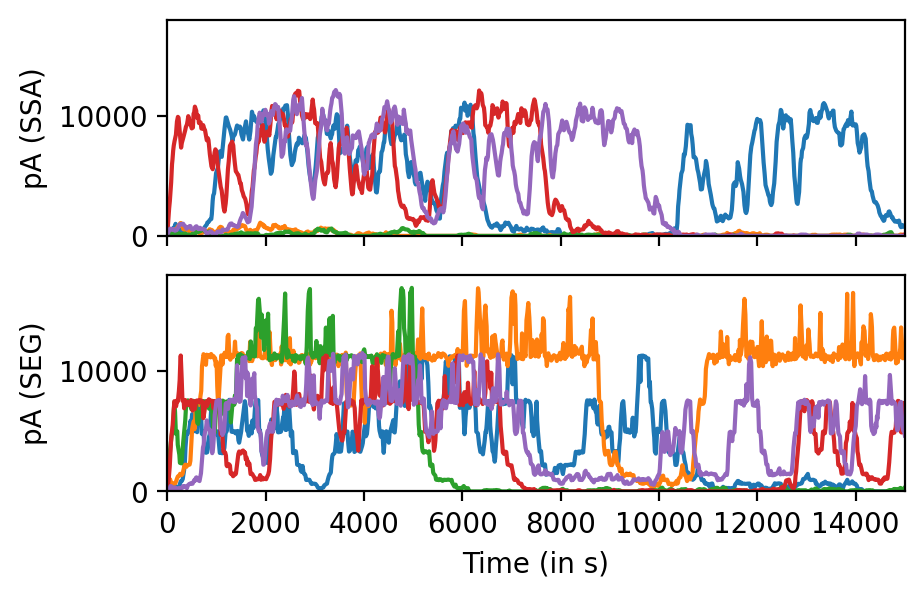}
    \caption{Visual comparison of simulations for the toggle switch model (left) two simulations with multiple species (right) five trajectories of pA species for SSA and segmental simulation with $\pc{=}1.5$ and $\pk{=}100$.}
    \label{fig:toggle_switch_sims}
\end{figure}

\begin{figure}[!h]
    \centering
    \includegraphics[width=0.45\linewidth]{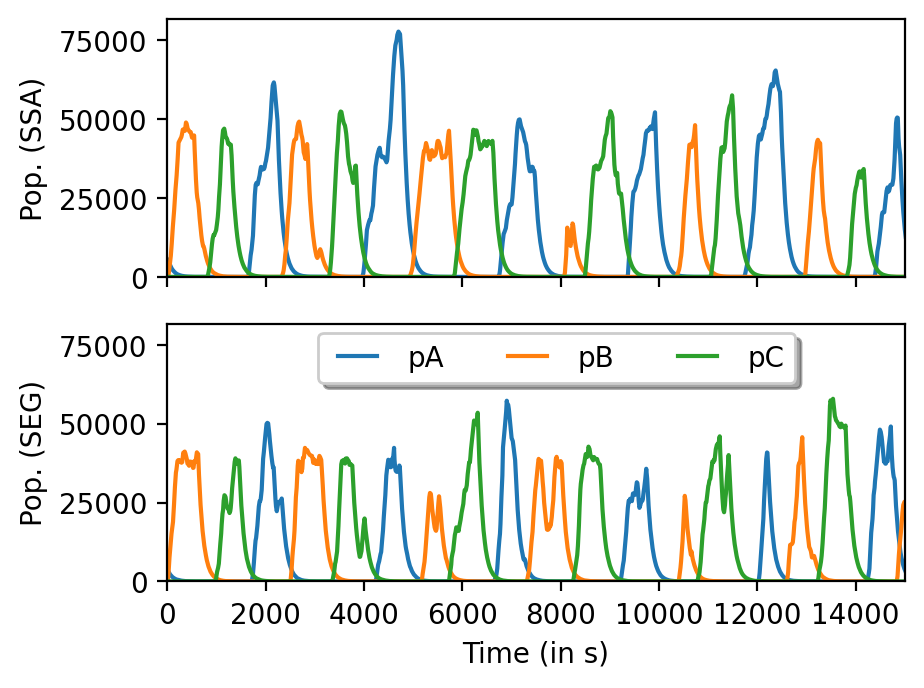}
    \includegraphics[width=0.45\linewidth]{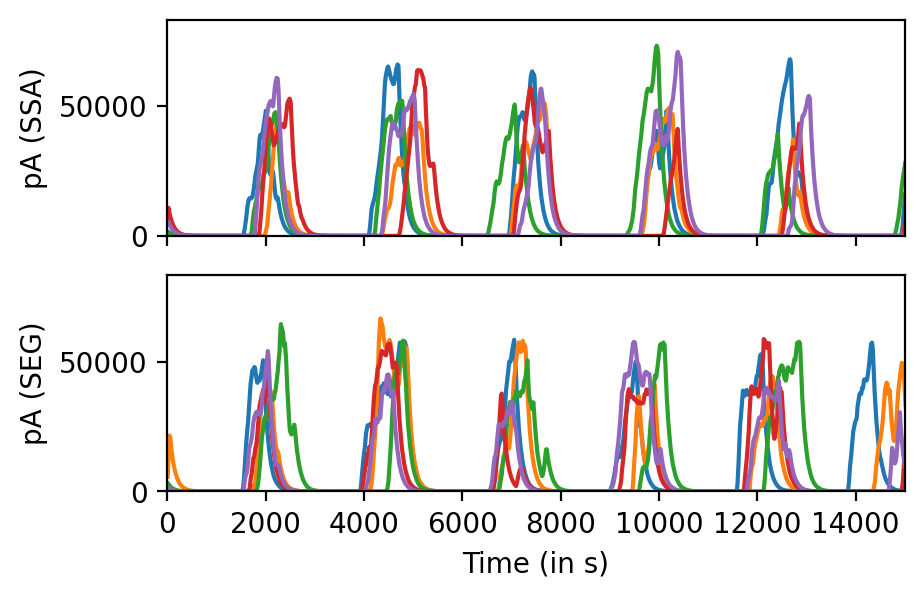}
    \caption{Visual comparison of simulations for the repressilator model (left) two simulations with multiple species (right) five trajectories of pA species for SSA and segmental simulation with $\pc{=}1.5$ and $\pk{=}100$.}
    \label{fig:repressilator_sims}
\end{figure}

\clearpage
\section{Quantitative comparison of segmental simulation} \label{appendix:quantitative}

\begin{figure}[!b]
    \vspace{-1.8cm}
    \centering
    \includegraphics[width=0.49\linewidth]{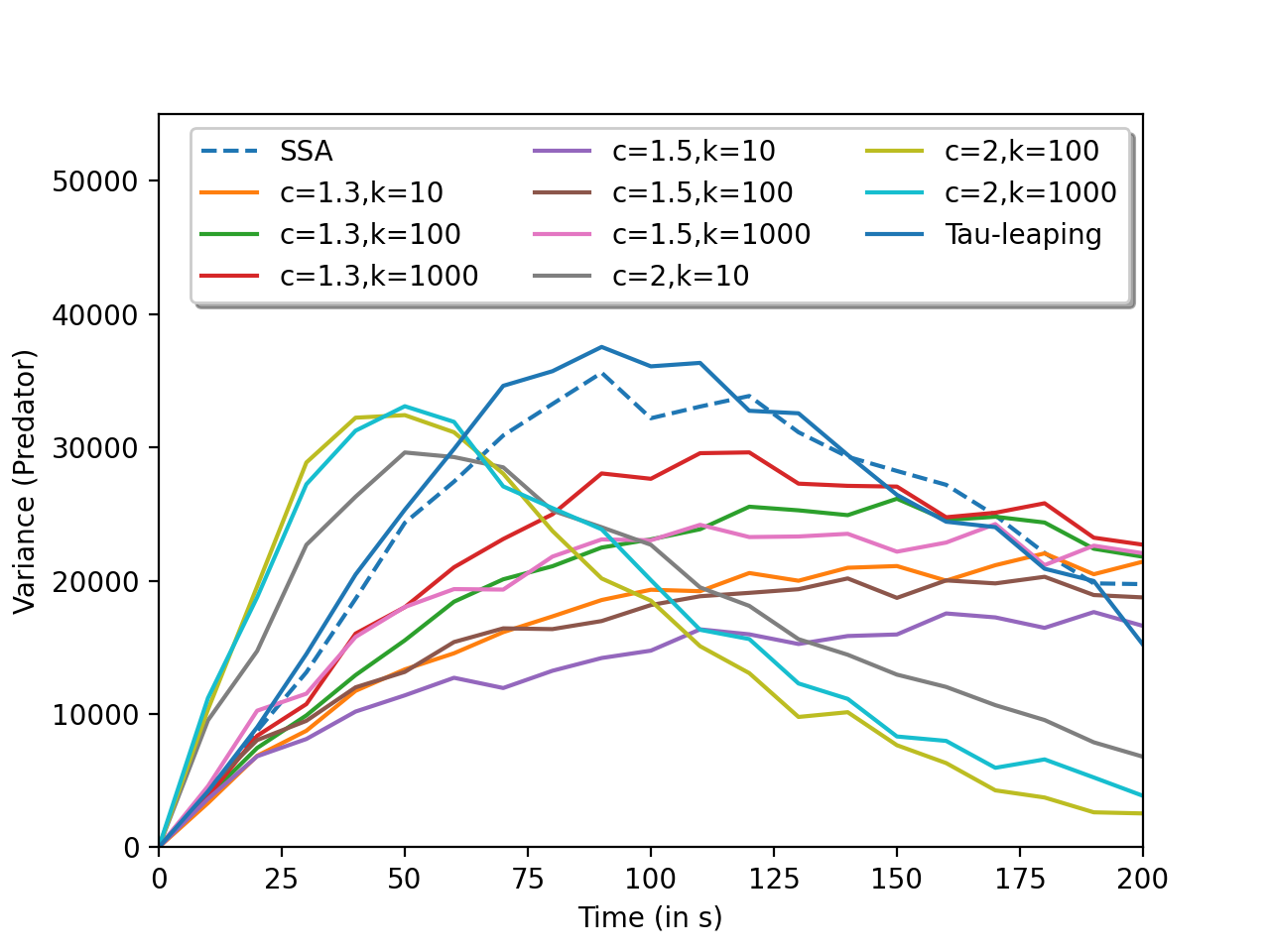}
    \includegraphics[width=0.49\linewidth]{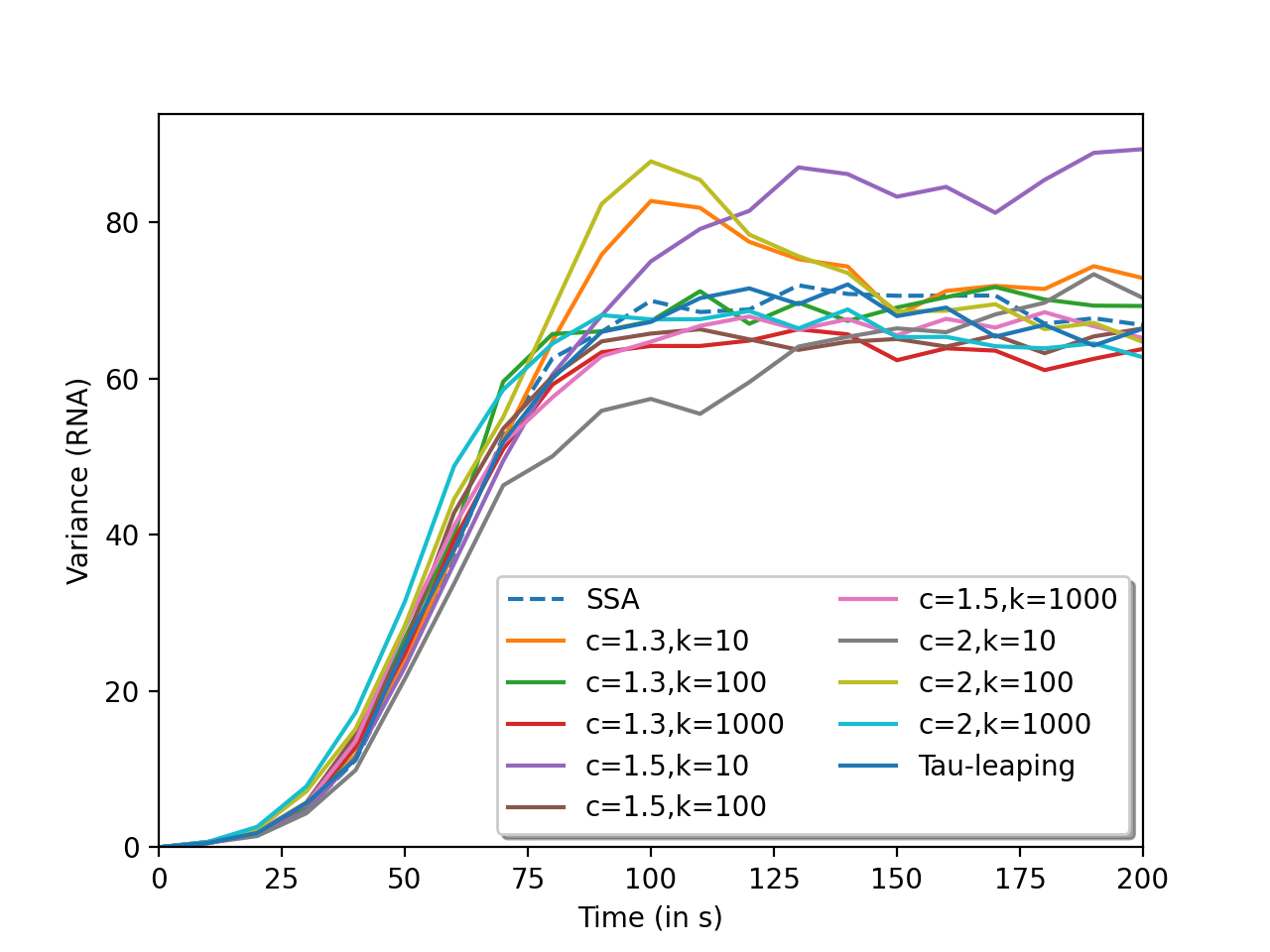} \\
    \includegraphics[width=0.49\linewidth]{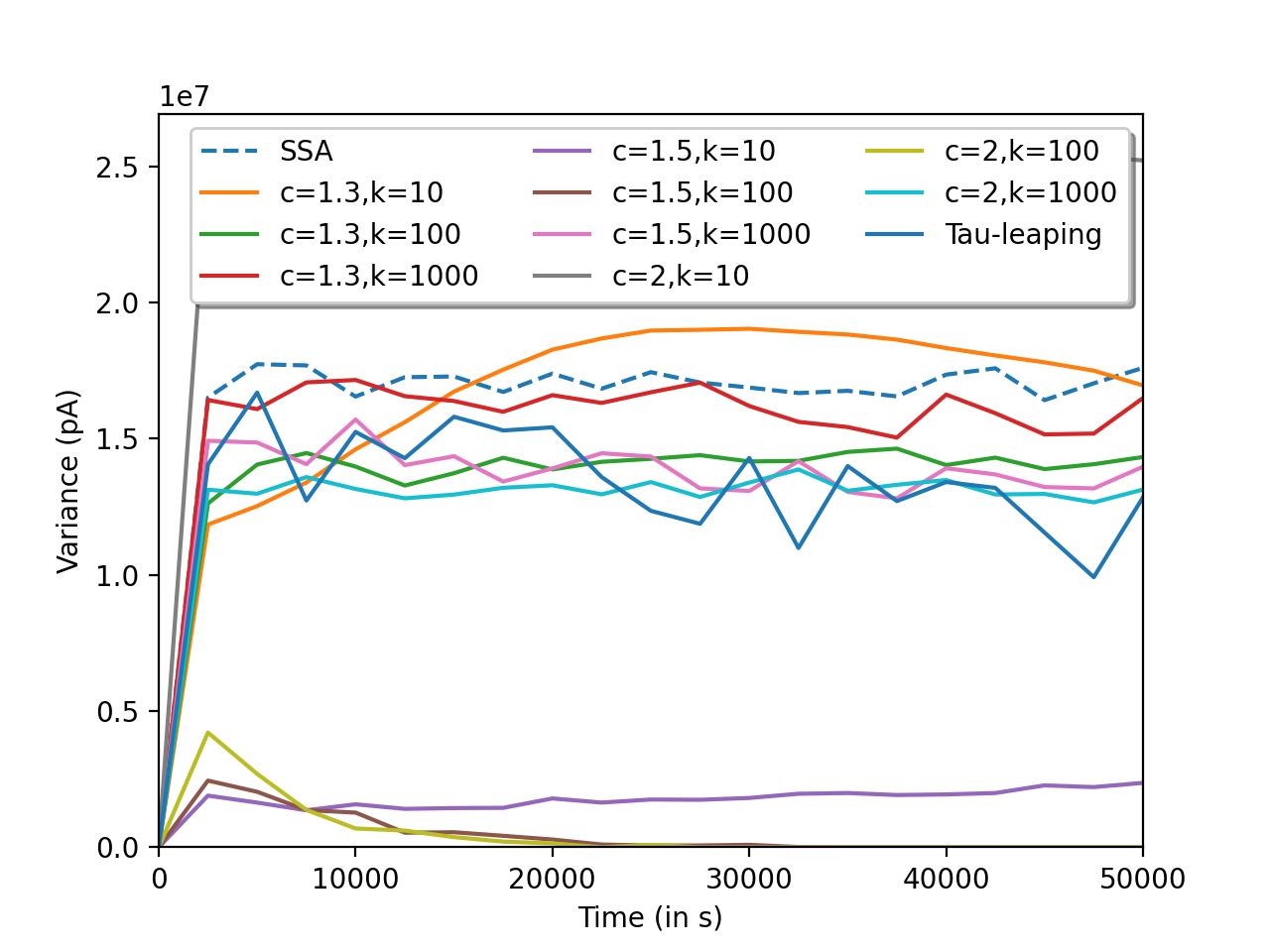}
    \includegraphics[width=0.49\linewidth]{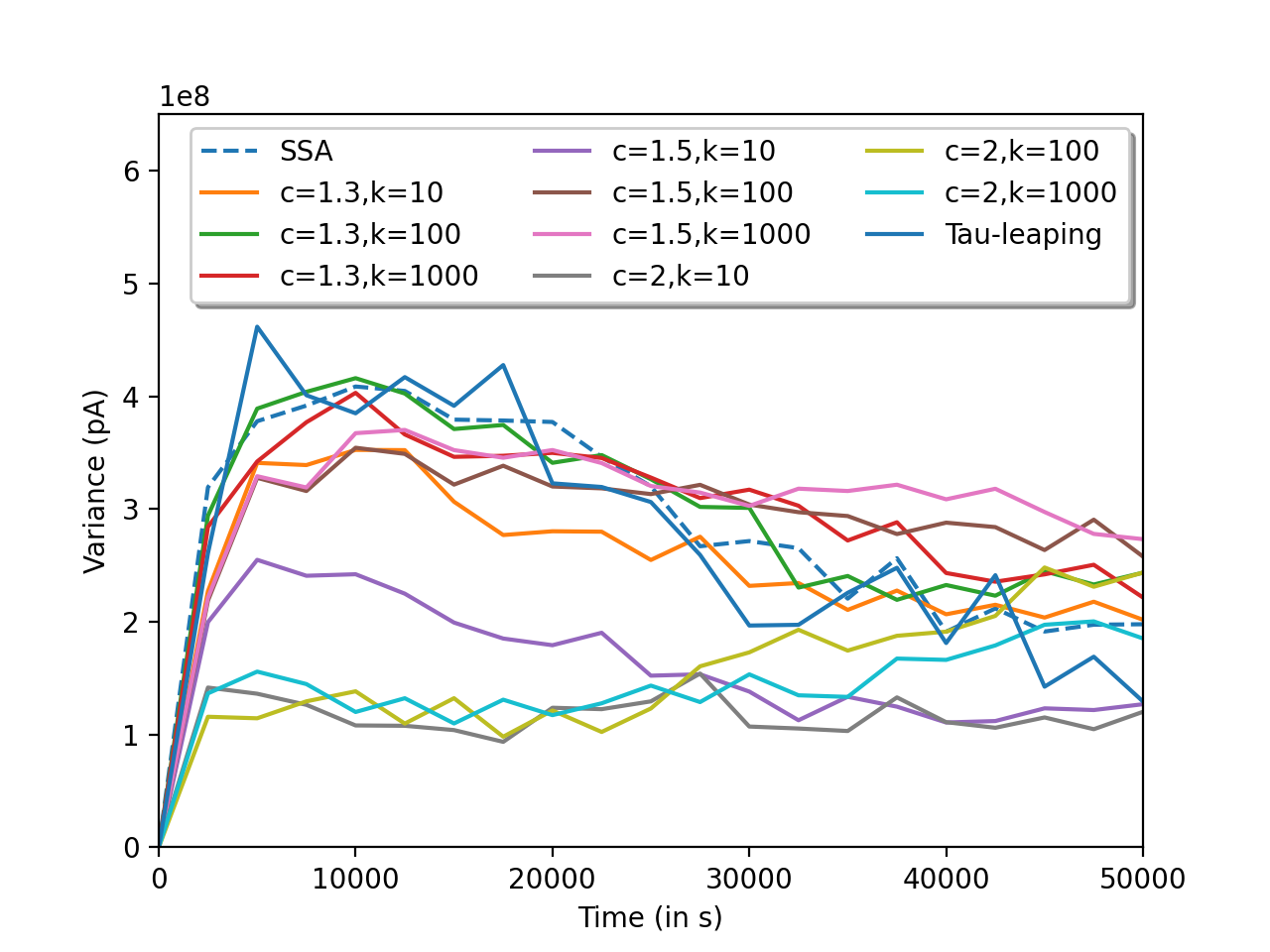}\\ \vspace{0.1cm}
    \caption{Variance over time for the models predator-prey (top left), viral infection (top right), toggle switch (bottom left) and repressilator (bottom right).}
    \label{fig:variance}
\end{figure}

\begin{table}[!b]
    \vspace{0.2cm}
    \centering
    \begin{tabular}{c||r||r|r|r||r|r|r||r|r|r|}
     \multirow{2}{*}{Model}  &     \multirow{2}{*}{SSA}        & \multicolumn{3}{c||}{SEG $k=10$} &  \multicolumn{3}{c||}{SEG $k=100$ } &  \multicolumn{3}{c|}{SEG $k=1000$ }\\ \cline{3-11}
     & & \multicolumn{1}{c|}{$c=2$} & \multicolumn{1}{c|}{$c=1.5$} & \multicolumn{1}{c||}{$c=1.3$} & \multicolumn{1}{c|}{$c=2$} & \multicolumn{1}{c|}{$c=1.5$} & \multicolumn{1}{c||}{$c=1.3$}  & \multicolumn{1}{c|}{$c=2$} & \multicolumn{1}{c|}{$c=1.5$} & \multicolumn{1}{c|}{$c=1.3$} \\ \hline \hline
PP & 8.9E4 & 700 & 760 & 2100 & 840 & 710 & 2800 & 810 & 710 & 2600\\ \hline
VI & 3.6E6 & 160 & 350 & 520 & 200 & 280 & 460 & 190 & 290 & 480\\ \hline
TS & 6.9E7 & 2.1E5 & 2.1E5 & 2.3E5 & 2.1E5 & 2.0E5 & 2.0E5 & 2.0E5 & 2.0E5 & 2.0E5\\ \hline
RP & 2.6E7 & 3.0E4 & 3.0E4 & 3.3E4 & 2.6E4 & 2.8E4 & 3.1E4 & 2.6E4 & 2.8E4 & 3.1E4\\ \hline
    \end{tabular}
    \vspace{1em}
    \caption{Average number of reactions per simulation for SSA and average number of applied summaries per simulation for different abstraction parameters.
    }
    \label{tab:parts}
\end{table}

\begin{table}[!b]
    \centering
    \begin{tabular}{c||r|r|r||r|r|r||r|r|r|}
     \multirow{2}{*}{Model} & \multicolumn{3}{c||}{SEG $k=10$} &  \multicolumn{3}{c||}{SEG $k=100$ } &  \multicolumn{3}{c|}{SEG $k=1000$ }\\ \cline{2-10}
     & \multicolumn{1}{c|}{$c=2$} & \multicolumn{1}{c|}{$c=1.5$} & \multicolumn{1}{c||}{$c=1.3$} & \multicolumn{1}{c|}{$c=2$} & \multicolumn{1}{c|}{$c=1.5$} & \multicolumn{1}{c||}{$c=1.3$}  & \multicolumn{1}{c|}{$c=2$} & \multicolumn{1}{c|}{$c=1.5$} & \multicolumn{1}{c|}{$c=1.3$} \\ \hline \hline
PP & 1.6E3 & 3.8E3 & 8.2E3 & 1.6E4 & 3.6E4 & 7.9E4 & 1.4E5 & 3.0E5 & 6.7E5\\ \hline
VI & 8.8E3 & 3.0E4 & 8.3E4 & 7.7E4 & 2.0E5 & 5.3E5 & 4.5E5 & 1.0E6 & 2.2E6\\ \hline
TS & 3.9E4 & 9.5E4 & 2.7E5 & 4.7E5 & 1.2E6 & 2.6E6 & 3.3E6 & 7.8E6 & 1.7E7\\ \hline
RP & 1.2E5 & 3.8E5 & 1.1E6 & 1.3E6 & 3.8E6 & 9.2E6 & 9.7E6 & 2.4E7 & 3.2E7\\ \hline
    \end{tabular}
    \vspace{1em}
    \caption{Number of saved summaries after 10,000 simulations for different
abstraction parameters.}
    \label{tab:summaries}
      \vspace{-0.8em}
\end{table}

\clearpage

\section{Benefits of working with concrete end points}\label{appendix:concrete}

Recall that the segmental simulation builds on the population level abstraction with accelerated transition~\cite{CAV19,CAV20} where the analysis exclusively works in the abstract domain, i.e. they only keep track of abstract states.
Intuitively, this means that we would simply ``jump'' to the representative of the current abstract state and apply the segment there. 
This kind of rounding is a major source of error and makes these kinds of abstractions fail even for some simple cases. 
The reason for this is that making significant progress in one dimension of the state space resets any progress made in all other dimensions.

\begin{figure}[!htb]
    \centering
    \begin{tikzpicture}[xshift=-100, thick, scale=0.5]
    \tikzstyle{abstractstate}=[black]
    \tikzstyle{state}=[circle, fill=gray, inner sep=0pt, minimum size=0.2cm]
    \tikzstyle{transition}=[very thick, ->]
    \tikzstyle{m1}=[purple]
    \tikzstyle{m2}=[blue]
    \tikzstyle{jump}=[dotted]
    
    % grid aka abstract states
    \foreach \i in {0,...,2}
    {
        \draw[abstractstate, dashed] (48,\i-0.5) -- (47.2,\i-0.5);
        \draw[abstractstate] (48,\i-0.5) -- (54,\i-0.5);
        \draw[abstractstate, dashed] (54,\i-0.5) -- (54.8,\i-0.5);
    }
    \draw[abstractstate] (48.5,-0.5) -- (48.5,1.5);
    \node at (46, 0) {OFF};
    \draw[abstractstate] (51.5,-0.5) -- (51.5,1.5);
    \node at (46, 1) {ON};
    \node at (46, 2) {\#X};
    \foreach \x in {48,...,54}
    {
        \node at (\x, 2) {\x};
        \foreach \y in {0,...,1}
        {
            \node[state] (\x_\y) at (\x, \y) {};
        }
    }
    \foreach \i in {50,52}
    {
        \pgfmathtruncatemacro{\ii}{(\i+1};
        \draw[transition, m1] (\i_1) -- (\ii_0);
    }
    \draw[transition, m1] (54_1) -- (54.8, 0.2);
    \foreach \i in {51,53}
    {
        \pgfmathtruncatemacro{\ii}{(\i+1};
        \draw[transition, m1] (\i_0) -- (\ii_1);
    }
    
    \begin{scope}[shift={(9,0)}]
        \foreach \i in {0,...,2}
        {
            \draw[abstractstate, dashed] (48,\i-0.5) -- (47.2,\i-0.5);
            \draw[abstractstate] (48,\i-0.5) -- (52,\i-0.5);
            \draw[abstractstate, dashed] (52,\i-0.5) -- (52.8,\i-0.5);
        }
        \draw[abstractstate] (48.5,-0.5) -- (48.5,1.5);
        \draw[abstractstate] (51.5,-0.5) -- (51.5,1.5);
        \foreach \x in {48,...,52}
        {
            \node at (\x, 2) {\x};
            \foreach \y in {0,...,1}
            {
                \node[state] (\x_\y) at (\x, \y) {};
            }
        }
        \draw[transition, m2] (50_1) -- (51_0);
        \draw[transition, m2] (50_0) -- (51_1);
        \draw[transition, m2, jump] (51_1) -- (50_1);
        \draw[transition, m2, jump] (51_0) -- (50_0);
    \end{scope}

    \end{tikzpicture}
    \vspace{0.5em}
    \caption{(left) Deterministic behavior of CRN with reactions $\textsc{ON} \rightarrow \textsc{OFF} + \textsc{X}$ and $\textsc{OFF} \rightarrow \textsc{ON} + \textsc{X}$ starting in initial state $(\textsc{ON},50\textsc{X})$ with growing number of $X$ molecules. (right) Incorrect trajectory of population abstraction using jumps that are depicted as dotted arrows.}
    \label{fig:rounding_is_bad}
\end{figure}
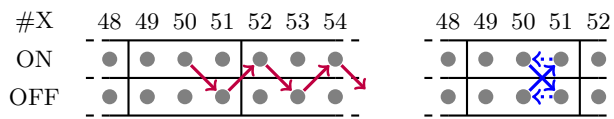

Fig.~\ref{fig:rounding_is_bad} illustrates this problem for a simple CRN with three species $\textsc{ON}$, $\textsc{OFF}$ and $\textsc{X}$, and two reactions $\textsc{ON} \rightarrow \textsc{OFF} + \textsc{X}$ and $\textsc{OFF} \rightarrow \textsc{ON} + \textsc{X}$. 
When starting in the state $(\textsc{ON}, 50\textsc{X})$ the system is deterministic and the number of $X$ molecules grows unbounded. 
Let us assume that the current abstract interval for $X$ is $[49{,}51]$ with representative $50$.
Then, with rounding, the abstraction incorrectly predicts that the number of $X$ molecules never exceeds $51$.
Intuitively, the constant on-off switching and the resulting rounding (depicted as dashed arrows), resets all progress in the dimension of $X$.
Note that this problem occurs for all (non-trivial) choices of population levels.

Segmental simulation does not face this issue because we apply the concrete effect of each segment to the current concrete state. I.e. there is no rounding and we correctly keep track of the growing number of $X$ molecules. 

\clearpage

\section{Models}\label{appendix:models}

We will now give the exact definitions of all models used in the evaluation. This includes their reactions together with their respective propensity functions, their initial state and the time horizon of interest.

\begin{table}[b!]

\setlength\tabcolsep{0.1cm}
\begin{tabularx}{\textwidth}{|l||l|X|}
	\hline
	\multicolumn{3}{|c|}{Predator Prey} \\
	\hline
	\hline
	\multicolumn{1}{|l||}{Species} & \multicolumn{2}{c|}{$\textsc{Pred}, \textsc{Prey}$} \\
	\hline
	\multicolumn{1}{|l||}{Initial state} & \multicolumn{2}{c|}{$\left( 200\times\textsc{Pred}, 200\times\textsc{Prey}\right) $} \\
	\hline
	\multicolumn{1}{|l||}{End time} & \multicolumn{2}{c|}{200s} \\
	\hline
	 Reactions & $rep: \textsc{Prey} \xrightarrow{1\cdot \textsc{Prey}} 2\textsc{Prey}$ & $eat: \textsc{Pred} + \textsc{Prey} \xrightarrow{0.005\cdot \textsc{Pred}\cdot \textsc{Prey}} 2\textsc{Pred}$\\
& $starve: \textsc{Pred} \xrightarrow{1\cdot \textsc{Pred}} \emptyset$ &\\
	\hline
\end{tabularx}

\vspace{0.3cm}

\begin{tabularx}{\textwidth}{|l||l|X|}
	\hline
	\multicolumn{3}{|c|}{Viral} \\
	\hline
	\hline
	\multicolumn{1}{|l||}{Species} & \multicolumn{2}{c|}{$\textsc{DNA}, \textsc{RNA}, \textsc{P}, \textsc{V}$} \\
	\hline
	\multicolumn{1}{|l||}{Initial state} & \multicolumn{2}{c|}{$\left(1\times\textsc{RNA}\right) $} \\
	\hline
	\multicolumn{1}{|l||}{End time} & \multicolumn{2}{c|}{200s} \\
	\hline
	 Reactions & $d0: \textsc{DNA} + \textsc{P} \xrightarrow{1.125E-5\cdot \textsc{DNA}\cdot \textsc{P}} \textsc{V}$ & $x: \textsc{RNA} \xrightarrow{1000\cdot \textsc{RNA}} \textsc{RNA} + \textsc{P}$\\
& $t: \textsc{DNA} \xrightarrow{0.025\cdot \textsc{DNA}} \textsc{DNA} + \textsc{RNA}$ & $p: \textsc{RNA} \xrightarrow{1\cdot \textsc{RNA}} \textsc{DNA} + \textsc{RNA}$\\
& $d2: \textsc{RNA} \xrightarrow{0.25\cdot \textsc{RNA}} \emptyset$ & $d5: \textsc{P} \xrightarrow{1.9985\cdot \textsc{P}} \emptyset$\\
	\hline
\end{tabularx}

\vspace{0.3cm}

\begin{tabularx}{\textwidth}{|l||l|l|X|}
	\hline
	\multicolumn{4}{|c|}{Toggle Switch} \\
	\hline
	\hline
	\multicolumn{1}{|l||}{Species} & \multicolumn{3}{c|}{$\textsc{mA}, \textsc{mB}, \textsc{sA}, \textsc{sB}, \textsc{pA}, \textsc{pB}$} \\
	\hline
	\multicolumn{1}{|l||}{Initial state} & \multicolumn{3}{c|}{$\emptyset $} \\
	\hline
	\multicolumn{1}{|l||}{End time} & \multicolumn{3}{c|}{50000s} \\
	\hline
	Reactions & $r0: \emptyset \xrightarrow{1} \textsc{mA}$ & $r5: \textsc{mA} \xrightarrow{5\cdot \textsc{mA}} \textsc{sA}$ & $r10: \textsc{sA} \xrightarrow{0.01\cdot \textsc{sA}} \emptyset$\\
	& $r1: \emptyset \xrightarrow{1} \textsc{mB}$ & $r6: \textsc{mB} \xrightarrow{5\cdot \textsc{mB}} \textsc{sB}$ & $r11: \textsc{sB} \xrightarrow{0.01\cdot \textsc{sB}} \emptyset$\\
	& $r2: \textsc{mA} \xrightarrow{0.1\cdot \textsc{mA}} \emptyset$ & $r7: \textsc{mB} + \textsc{sA} \xrightarrow{20\cdot \textsc{mB}\cdot \textsc{sA}} \textsc{sA}$ & $r12: \textsc{sA} \xrightarrow{10\cdot \textsc{sA}} \textsc{sA} + \textsc{pA}$\\
	& $r3: \textsc{mB} \xrightarrow{0.1\cdot \textsc{mB}} \emptyset$ & $r8: \textsc{mA} + \textsc{sB} \xrightarrow{20\cdot \textsc{mA}\cdot \textsc{sB}} \textsc{sB}$ & $r13: \textsc{sB} \xrightarrow{10\cdot \textsc{sB}} \textsc{sB} + \textsc{pB}$\\
	& $r4: \textsc{pA} \xrightarrow{0.1\cdot \textsc{pA}} \emptyset$ & $r9: \textsc{pB} \xrightarrow{0.1\cdot \textsc{pB}} \emptyset$ &\\
	\hline
\end{tabularx}

\vspace{0.3cm}

\begin{tabularx}{\textwidth}{|l||l|X|}
	\hline
	\multicolumn{3}{|c|}{Repressilator} \\
	\hline
	\hline
	\multicolumn{1}{|l||}{Species} & \multicolumn{2}{c|}{$\textsc{mA}, \textsc{mB}, \textsc{mC}, \textsc{pA}, \textsc{pB}, \textsc{pC}$} \\
	\hline
	\multicolumn{1}{|l||}{Initial state} & \multicolumn{2}{c|}{$\left( 10\times\textsc{mA}, 500\times\textsc{pA} \right) $} \\
	\hline
	\multicolumn{1}{|l||}{End time} & \multicolumn{2}{c|}{50000s} \\
	\hline
	 Reactions & $spawnA: \emptyset \xrightarrow{0.1} \textsc{mA}$ & $despawnC: \textsc{mC} \xrightarrow{0.01\cdot \textsc{mC}} \emptyset$\\
	& $spawnB: \emptyset \xrightarrow{0.1} \textsc{mB}$ & $degradeA: \textsc{mA} + \textsc{pB} \xrightarrow{50\cdot \textsc{mA}\cdot \textsc{pB}} \textsc{pB}$\\
	& $spawnC: \emptyset \xrightarrow{0.1} \textsc{mC}$ & $degradeB: \textsc{mB} + \textsc{pC} \xrightarrow{50\cdot \textsc{mB}\cdot \textsc{pC}} \textsc{pC}$\\
	& $prodA: \textsc{mA} \xrightarrow{50\cdot \textsc{mA}} \textsc{mA} + \textsc{pA}$ & $degradeC: \textsc{mC} + \textsc{pA} \xrightarrow{50\cdot \textsc{mC}\cdot \textsc{pA}} \textsc{pA}$\\
	& $prodB: \textsc{mB} \xrightarrow{50\cdot \textsc{mB}} \textsc{mB} + \textsc{pB}$ & $dissolveA: \textsc{pA} \xrightarrow{0.01\cdot \textsc{pA}} \emptyset$\\
	& $prodC: \textsc{mC} \xrightarrow{50\cdot \textsc{mC}} \textsc{mC} + \textsc{pC}$ & $dissolveB: \textsc{pB} \xrightarrow{0.01\cdot \textsc{pB}} \emptyset$\\
	& $despawnA: \textsc{mA} \xrightarrow{0.01\cdot \textsc{mA}} \emptyset$ & $dissolveC: \textsc{pC} \xrightarrow{0.01\cdot \textsc{pC}} \emptyset$  \\           
	& $despawnB: \textsc{mB} \xrightarrow{0.01\cdot \textsc{mB}} \emptyset$ & \\
	\hline
\end{tabularx}
\end{table}

\end{document}